\documentclass{emulateapj}
\usepackage{apjfonts}
\usepackage{natbib}

\newcommand{\chandra}{{\it Chandra\/}}
\def\asca{{\it ASCA\/}}
\newcommand{\flux}{{erg~cm$^{-2}$~s$^{-1}$}}
\newcommand{\lum}{{erg~s$^{-1}$}}

\begin{document}

\title{Probing the X-Ray Binary Populations of the Ring Galaxy NGC 1291}

\author{
B.~Luo,\altaffilmark{1}
G.~Fabbiano,\altaffilmark{1}
T.~Fragos,\altaffilmark{1}
D.-W.~Kim,\altaffilmark{1}
K.~Belczynski,\altaffilmark{2,3}
N.~J.~Brassington,\altaffilmark{4}
S.~Pellegrini,\altaffilmark{5}
P.~Tzanavaris,\altaffilmark{6,7,8}
J.~Wang,\altaffilmark{1}
and A.~Zezas\altaffilmark{9}
}
\altaffiltext{1}{
Harvard-Smithsonian Center for Astrophysics,
60 Garden Street, Cambridge, MA 02138, USA}
\altaffiltext{2}{
Astronomical Observatory, University of Warsaw, Al. Ujazdowskie 4,
00-478 Warsaw, Poland}
\altaffiltext{3}{
Center for Gravitational Wave Astronomy, University of Texas at
Brownsville, Brownsville, TX 78520, USA}
\altaffiltext{4}{
School of Physics, Astronomy and Mathematics, University of Hertfordshire, College Lane, Hatfield AL10 9AB, UK}
\altaffiltext{5}{
Dipartimento di Astronomia, Universit\'a di Bologna, Via Ranzani 1,
40127 Bologna, Italy}
\altaffiltext{6}{Laboratory for X-ray Astrophysics, NASA Goddard Space
Flight Center, Greenbelt, MD 20771, USA}
\altaffiltext{7}{Department of Physics and Astronomy, The Johns Hopkins
University, Baltimore, MD 21218, USA}
\altaffiltext{8}{NASA Post-doctoral Program Fellow}
\altaffiltext{9}{
Physics Department, University of Crete, P.O. Box 2208, GR-710 03, Heraklion, Crete, Greece}
\begin{abstract}
We present \chandra\ studies of the X-ray binary (XRB) populations in 
the bulge and ring regions of 
the ring galaxy NGC 1291. We detect 169 X-ray point sources in the 
galaxy, 75 in the bulge and 71 in the ring, utilizing the
four available \chandra\ observations totaling an effective 
exposure of 179 ks.
We report photometric properties of these sources in a point-source catalog.
There are $\approx40\%$ of the bulge sources and
$\approx25\%$ of the ring sources showing $>3\sigma$ long-term variability in
their X-ray count rate. The X-ray colors suggest that
a significant
fraction of the bulge ($\approx75\%$) and ring ($\approx65\%$)
sources are likely low-mass X-ray binaries (LMXBs). 
The spectra of the nuclear source indicate that it is
a low-luminosity AGN with moderate obscuration; spectral variability is
observed between individual observations.
We construct 0.3--8.0 keV X-ray luminosity functions (XLFs)
for the bulge and ring XRB populations, taking into
account the detection incompleteness and background AGN
contamination. We reach 90\% completeness limits 
of $\approx1.5\times10^{37}$ and
$\approx2.2\times10^{37}$~\lum\ for the bulge and ring populations,
respectively. Both XLFs can be fit with a broken power-law model, and
the shapes are consistent with those expected for populations dominated by
LMXBs. 
We perform detailed population synthesis modeling of the 
XRB populations in NGC 1291, which 
suggests that 
the observed combined XLF is dominated by an old LMXB population.
We compare the bulge and ring XRB populations, and 
argue that the ring XRBs are associated with a younger 
stellar population than the bulge sources,
based on the relative overdensity of X-ray sources in the ring, 
the generally harder X-ray color of the ring sources, 
the overabundance of luminous sources in the combined XLF,
and the flatter shape of the ring XLF.
\end{abstract}

\section{INTRODUCTION}

Galactic rings are a peculiar phenomenon among galaxies. A prototype 
ring galaxy is the famous Cartwheel galaxy, with a spectacular ring luminous
in the infrared (IR), optical, UV, and X-ray bands.
The formation of ring-like structures in galaxies is not fully understood. 
The two most common theories are the collisional origin and the
resonance origin.
A collisional ring could be created by
a head-on collision between a compact 
companion galaxy and a larger disk system; a collisional ring galaxy usually
contains an off-centered nucleus and 
an asymmetric ring structure, and there is a significant chance of finding 
a companion galaxy close to the ring galaxy (see \citealt{Appleton1996}
for a review). A resonance ring could be produced by the 
outer Lindblad resonance of spiral bars; such a system is preferentially
observed in barred galaxies, and the resulting ring is generally symmetric
(see \citealt{Buta1996} for a review). 
There are a few systems that are difficult to
explain with these two theories, such as the Hoag's Object \citep{Hoag1950},
for which alternative formation scenarios have been proposed 
\citep[e.g.,][]{Finkelman2011}.

Galactic rings are expected to exhibit significant star formation compared
to their progenitor galaxies mainly due to the 
enhanced gas densities in the rings,
regardless of the formation mechanism. Surveys of ring galaxies
in optical, H$\alpha$, IR, and radio 
have revealed that the rings are generally rich in H{\sc i} gas and 
H{\sc ii} regions \citep[e.g.,][]{van1988,Marston1995,Phillips1996,Ryder1996,
Grouchy2010}.
With \chandra's subarcsecond angular resolution, it is now possible
to study in detail the X-ray binary (XRB) populations in nearby galaxies,
e.g., low-mass X-ray binaries (LMXBs) that are associated with 
old stellar populations and high-mass X-ray 
binaries (HMXBs) that are associated with young stellar populations 
(see e.g., \citealt{Fabbiano2006} for a review).
Therefore, X-ray observations have become an independent 
approach for constraining
the nature of the underlying stellar populations by means of 
X-ray luminosity functions (XLFs) and X-ray colors of the XRBs.
However, little is known regarding the X-ray properties of galactic rings, 
except for the Cartwheel galaxy, 
which has an exceptionally large number of young ($\le 10^7$~yr)
ultraluminous X-ray sources (ULXs; $L_{\rm X}\ga10^{39}$~\lum;
e.g., \citealt{Swartz2004}) in the star-forming
ring (e.g., \citealt{Gao2003,King2004,Wolter2004}).

\begin{figure*}
\centerline{
\includegraphics[scale=1.]{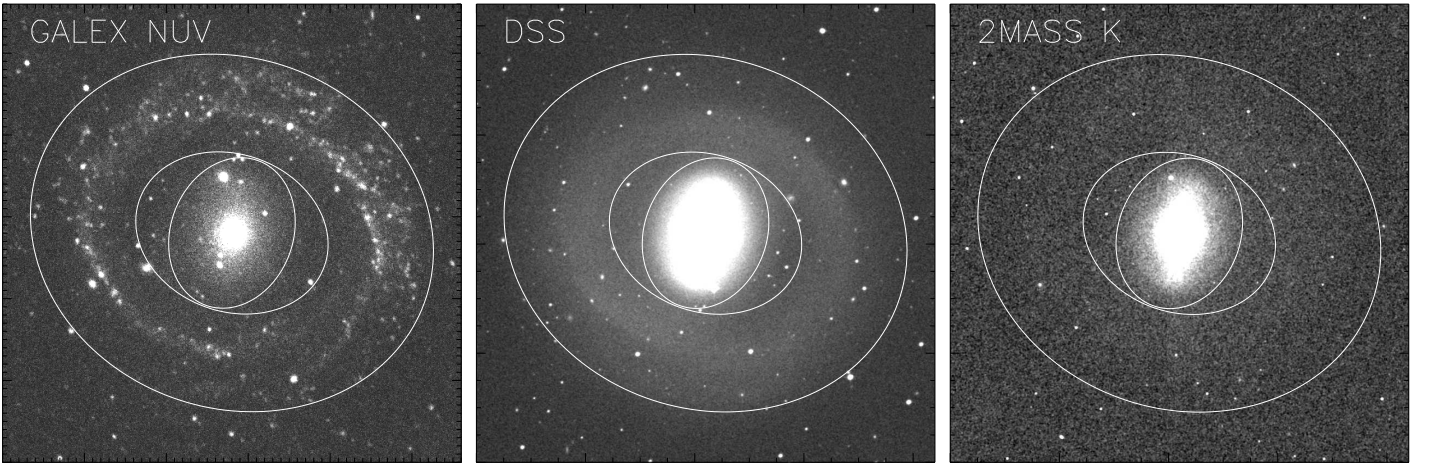}
}
\figcaption{
Images of NGC 1291 in the {\it GALEX} near-UV (NUV, $\lambda_{\rm eff}=2267$~\AA), Digitized Sky Survey 
red (RG610),
and 2MASS $K$ bands. 
The size of each image is $14\arcmin\times14\arcmin$, centered on the galactic
nucleus.
The ellipses outline the bulge 
and ring regions (see Section 3 for details). The UV ring appears to be dominated by young stars.
\label{img}}
\end{figure*}

NGC 1291 is a remarkable ring galaxy, 
first noted by \citet{Perrine1922}.
It was classified as (R)SB(s)0/a (RC3; \citealt{deVaucouleurs1991}),
and is virtually face-on \citep{Mebold1979}.
The images of NGC 1291 in the UV, optical, and near-IR (NIR) are shown in 
Figure \ref{img}.
The galaxy includes a fainter outer ring
and a prominent bulge, which 
was considered as a prototypical pseudobulge \citep[e.g.,][]{Kormendy2004}.
At a distance of 8.9 Mpc, NGC 1291 has $M_{\rm B,0}$=$-20.6$ 
\citep{deVaucouleurs1975}.
The D$_{25}$ ellipse of NGC 1291 has a size of
$9.8\arcmin\times8.1\arcmin$ \citep{Kennicutt2003},
and the nuclear position of the galaxy is 
$\alpha_{\rm J2000.0}=
03^{\rm h}17^{\rm m}18.6^{\rm s}$, $\delta_{\rm J2000.0}=-41\degr06\arcmin29\arcsec$ \citep{Evans2010}. 
The Galactic column density along the line of sight to NGC 1291
is \hbox{$N_{\rm H}=2.1\times 10^{20}$~cm$^{-2}$} \citep{Dickey1990}.\footnote{See 
{http://heasarc.nasa.gov/cgi-bin/Tools/w3nh/w3nh.pl}.}

The ring of NGC 1291 does not appear as energetic as that of the 
Cartwheel galaxy. There were only two ULXs reported in the ring 
based on the first \chandra\ observation in 2000 \citep{Swartz2004}. 
The total star formation rate (SFR) 
of NGC 1291 is only 0.4~$M_{\sun}$~yr$^{-1}$ 
\citep{Kennicutt2003}, 
compared
to that of 67.5~$M_{\sun}$~yr$^{-1}$ for the Cartwheel galaxy \citep{Higdon1993}.
Given its barred spiral nature, the relatively
symmetric structure of the ring, and the absence
of any apparent companion, NGC 1291 is likely a resonance ring system,
while the Cartwheel galaxy is most likely a
collisional ring system \citep[e.g.,][]{Appleton1996}.
Intense star formation still appears present in the ring of NGC 1291,
which is obvious from the UV image (Figure \ref{img}).
On the contrary, the bulge is much redder than the ring, and dominates
in the NIR image (Figure \ref{img}).
\hbox{\citet{van1988}} mapped the H{\sc i} distribution in NGC 1291, 
and revealed that H{\sc i} gas is concentrated in the ring region.
The ring also contains discrete H{\sc ii} regions, while the bulge has
filaments of ionized gas \citep{Crocker1996}.
Given the UV/optical/radio data, the ring appears to host 
a younger stellar population than the bulge.
\citet{Noll2009} studied the star-formation history of many galaxies
including NGC 1291, using spectral energy distribution fitting
techniques.
The best-fit model for NGC 1291 (Noll, private communication) includes
two exponential starburst events: one happened 10~Gyr ago with
an e-folding timescale of 0.25~Gyr and the other happened 200~Myr ago with
an e-folding timescale of 50~Myr.
The stellar mass of the old population is $\approx 330$ times higher
than that of the younger one. Although the fitting results were derived 
for the entire galaxy and may have large uncertainties, they are consistent
with the UV/optical/radio data if the ring hosts the younger 
stellar population.

NGC 1291 had been observed with \chandra\ twice in June and November 2000,
with a total exposure of $\approx 80$~ks.
A first study by \citet{Irwin2002} was limited to the bulge region.
\citet{Kilgard2005} presented a
point-source catalog based on these observations, and found that 
the XLF and X-ray colors
suggest old XRB populations in the galaxy. This study however,
did not discriminate
between bulge and ring XRBs.

We obtained two additional \chandra\ observations of NGC 1291 in May 2010, 
with a total exposure of $\approx 120$~ks. The new observations cover
the majority of the ring region, allowing us to investigate the XRB
populations in the bulge and ring respectively. NGC 1291 is much 
closer than the Cartwheel galaxy 
($\approx10$~Mpc compared to $\approx120$~Mpc), and thus we are able to
probe the ring XRB population down to an unprecedented
luminosity limit ($\approx10^{37}$~\lum). We compare the ring XRB 
population to the bulge population and test whether 
the galactic ring hosts a younger stellar population from 
an X-ray point of view. The deep observations on the bulge also allow us
to compare its LMXB population to those of the recently 
acquired deep exposures of elliptical galaxies, and establish if all LMXB 
populations are indeed compatible.

This paper is organized as follows.
We describe
the \chandra\ observations of NGC 1291 and the
data reduction procedure in Section 2.
In Section 3 we construct an X-ray point-source catalog. 
X-ray source properties are presented in Section 4, including 
X-ray source radial profile, source variability, and X-ray colors. 
The nuclear source and ULX candidates are also discussed.
In Section 5, we calculate and parameterize 
the XLFs for the bulge and ring regions, respectively,
taking into account the detection incompleteness and 
background AGN contamination.
We model the observed XLF using population synthesis (PS) models in Section 6.
In Section 7, we discuss implications of the X-ray properties and 
we argue that the ring hosts a younger stellar population than the bulge.
We
summarize our results in Section 8.

\section{OBSERVATIONS AND DATA REDUCTION}\label{datared}
NGC 1291 has been observed with the S3 chip of \chandra\ Advanced CCD Imaging
Spectrometer (ACIS; \citealt{Garmire2003}) at four epochs.
The four observations are listed in Table~\ref{tbl-obs} along with their exposure times,
ranging from 37~ks to 70~ks. 
The focal-plane
temperature was $-120\degr$C for all the observations.
We reduced and analyzed the observational data 
using mainly the \chandra\ Interactive Analysis
of Observations (CIAO) tools. \footnote{See
http://cxc.harvard.edu/ciao/ for details on CIAO.}
We used the {\sc chandra\_repro} script to reprocess the
data with the latest calibration.
The background light curve of each observation was then inspected
and background flares were removed using the {\sc deflare}
CIAO script. The flare-cleaned exposure times are also listed in Table~\ref{tbl-obs};
there was a significant flare in observation 2059 that reduces its usable exposure
by $\approx40\%$. The total usable exposure for NGC~1291 is 178.6 ks.

We registered the astrometry of the observations using the CIAO tool
{\sc reproject\_aspect}.
We created a 0.3--8.0 keV image for each observation and searched for sources
using {\sc wavdetect} \citep{Freeman2002}
at a false-positive probability threshold of 10$^{-6}$. These source lists
were used to register the observations to the astrometric frame of observation
11272 with the CIAO script {\sc reproject\_aspect}, 
adopting a 3$\arcsec$ matching radius and a
residual rejection limit of 0.6$\arcsec$.
We reprojected the registered event lists to observation
11272 using {\sc reproject\_events}, and then merged all the observations
to create a master event file using {\sc dmmerge}.
The ACIS-S3 chip has different pointings for the four observations, and
the average aim point (weighted by exposure time) is
$\alpha_{\rm J2000.0}=
03^{\rm h}17^{\rm m}09.31^{\rm s}$, $\delta_{\rm J2000.0}=-41\degr06\arcmin25.1\arcsec$.

We constructed images from the merged event file using the standard
\asca\ grade set (\asca\ grades 0, 2, 3, 4, 6) for six bands:
\hbox{0.3--8.0~keV} (full band), \hbox{0.3--2.0~keV} (soft band), 
\hbox{0.5--2.0~keV} (conventional soft band),
\hbox{2.0--8.0~keV} (hard band), \hbox{0.3--1.0~keV} (soft band 1), 
and \hbox{1.0--2.0~keV} (soft band 2).
For each observation, we created exposure maps in these bands
following the basic procedure
outlined in Section 3.2 of \citet{Hornschemeier2001}, which
takes into account the effects of vignetting, gaps between the CCDs,
bad-column filtering, bad-pixel filtering, and the spatially
dependent degradation in quantum efficiency
due to contamination on the ACIS optical-blocking filters.
A photon index of
$\Gamma=1.7$ was assumed in creating the exposure map,
which is a typical value for photon indices of X-ray binaries 
\citep[e.g.,][]{Irwin2003,Brassington2010}. 
Merged exposure maps were then created from
the exposure maps of the individual observations.

\begin{figure}
\centerline{
\includegraphics[scale=0.5]{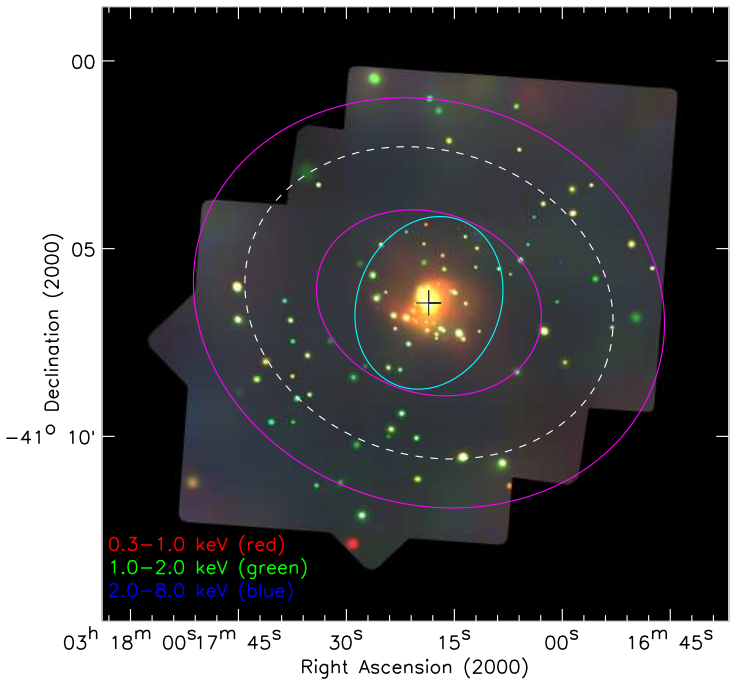}
}
\figcaption{
\chandra\ ``false-color'' image of NGC 1291. This image
is a color composite of the  exposure-corrected adaptively smoothed images
in the 0.3--1.0 keV ({\it red}), 1.0--2.0 keV ({\it green}), and 2.0--8.0
keV ({\it blue}) bands.
The cross symbol shows the center of the galaxy.
The white dashed ellipse shows the D$_{25}$ region \citep{Kennicutt2003}.
The cyan and magenta ellipses outline the bulge and ring regions of the galaxy,
respectively.
\label{clrimg}}
\end{figure}

We created adaptively smoothed images from the raw images using the CIAO
tool {\sc csmooth}.
Exposure-corrected smoothed images were then constructed following
Section 3.3 of \citet{Baganoff2003}.
We show in
Figure~\ref{clrimg} a color composite of the exposure-corrected
smoothed images
in the 0.3--1.0, 1.0--2.0, and 2.0--8.0
keV bands.

\section{POINT-SOURCE CATALOG} \label{catalog}

We ran {\sc wavdetect} on the full-band
image \footnote{Source searching
was performed only in the full band as we focused on the study
of the full-band XLFs.}
to search for X-ray point sources, with a
``$\sqrt{2}$~sequence'' of wavelet scales (i.e.,\ 1, $\sqrt{2}$, 2,
$2\sqrt{2}$, 4, $4\sqrt{2}$, 8, $8\sqrt{2}$, and 16 pixels)
and a false-positive probability threshold of 10$^{-6}$.
At this threshold, we expect approximately two false detections given the size
of the image \citep[e.g.,][]{Kim2004}.

There are 169 X-ray sources detected in the full band.
We utilized the {\sc acis extract} (AE; \citealt{Broos2010}) program to
derive the source photometry. The main advantage of AE compared to
a simple circular-aperture extraction approach is that
AE uses a polygonal extraction region to represent
the point spread function (PSF)
for each observation of each source individually,
which is more suitable for multiple observations
with different aim points and roll angles.
Briefly, AE constructs source extraction regions that approximate
$\approx90\%$ encircled-energy fraction (EEF) contours of the PSFs
at 1.50 keV for each exposure. 
Smaller extraction regions (as low as $\approx60\%$)
were used for sources in crowded regions.
Background counts were extracted in an annular region around each source
excluding any pixels that belong to detected sources;
the number of extracted background counts was required to be at least
100 to reduce the uncertainty from the background estimation.
Encircled-energy fractions at several energies
(ranging from 0.28 to 8.60 keV) were provided by AE, and thus
we derived aperture corrections for all the six X-ray bands listed in 
Section \ref{datared} via interpolation.
Aperture-corrected net (background-subtracted) source counts
and the corresponding
1$\sigma$ errors \citep{Gehrels1986} for these bands were then computed;
the errors were propagated through the errors of the extracted sources and 
background counts following the numerical
method described in \S1.7.3 of \citet{Lyons1991}.
As the source detection was performed only in the full band,
we consider a source as undetected in a given band 
(other than the full band)
if it is less significant than 3$\sigma$, when calculated
using the net counts and 1$\sigma$ error in that band.
An upper limit on the source counts was derived for an undetected source,
using the Bayesian approach of \citet{Kraft1991} for the 99\% confidence level
if the extracted number of counts is less than 10 in the given band; otherwise, 
we calculated a 3$\sigma$ upper limit following the Poisson 
statistic \citep{Gehrels1986}.

An X-ray spectrum for each source was extracted by AE along with the spectrum
of its local background. We performed spectral fitting of each source
using XSPEC (Version 12.6.0; \citealt{Arnaud1996}) with the
Cash fitting statistic (CSTAT; \citealt{Cash1979}).
An
absorbed power-law model ({\sc tbabs*pow}) was employed for 
sources with more than 30 full-band counts; for the other sources, 
the spectra were modeled with a simple power law with the photon index fixed 
at $\Gamma=1.7$. The observed full-band fluxes were derived from
the best-fit spectra, which were then converted to full-band luminosities.
We applied a smaller correction factor (1.05) to the observed full-band
luminosities to
account for the Galactic absorption.

Of the 169 point sources, 75 are in the bulge region and 71 in the ring area.
We defined the bulge and ring regions based on the UV and NIR images shown in
Figure 1.
We examined the surface brightness contours of the $K$-band image and
selected a contour that encloses $\approx80\%$ of the light within the 
D$_{25}$ ellipse. This contour can be well approximated by an 
$4.7\arcmin\times3.7\arcmin$ ellipse centered on the galactic 
nucleus, which was used to defined the
bulge boundary. The ring region is defined by two ellipses centered 
on the galactic nucleus.
We fixed the position angle (70\degr) and axis ratio (1.15) of the 
ellipses according to the morphology information given in 
\citet{deVaucouleurs1975} and only allowed the
sizes of the ellipses to vary so that the ring region includes the majority of 
the prominent UV ring features. We note that $\approx85\%$ of ring region
is covered by the \chandra\ exposure (see Figure~\ref{clrimg}).

\begin{figure*}
\centerline{
\includegraphics[scale=0.5]{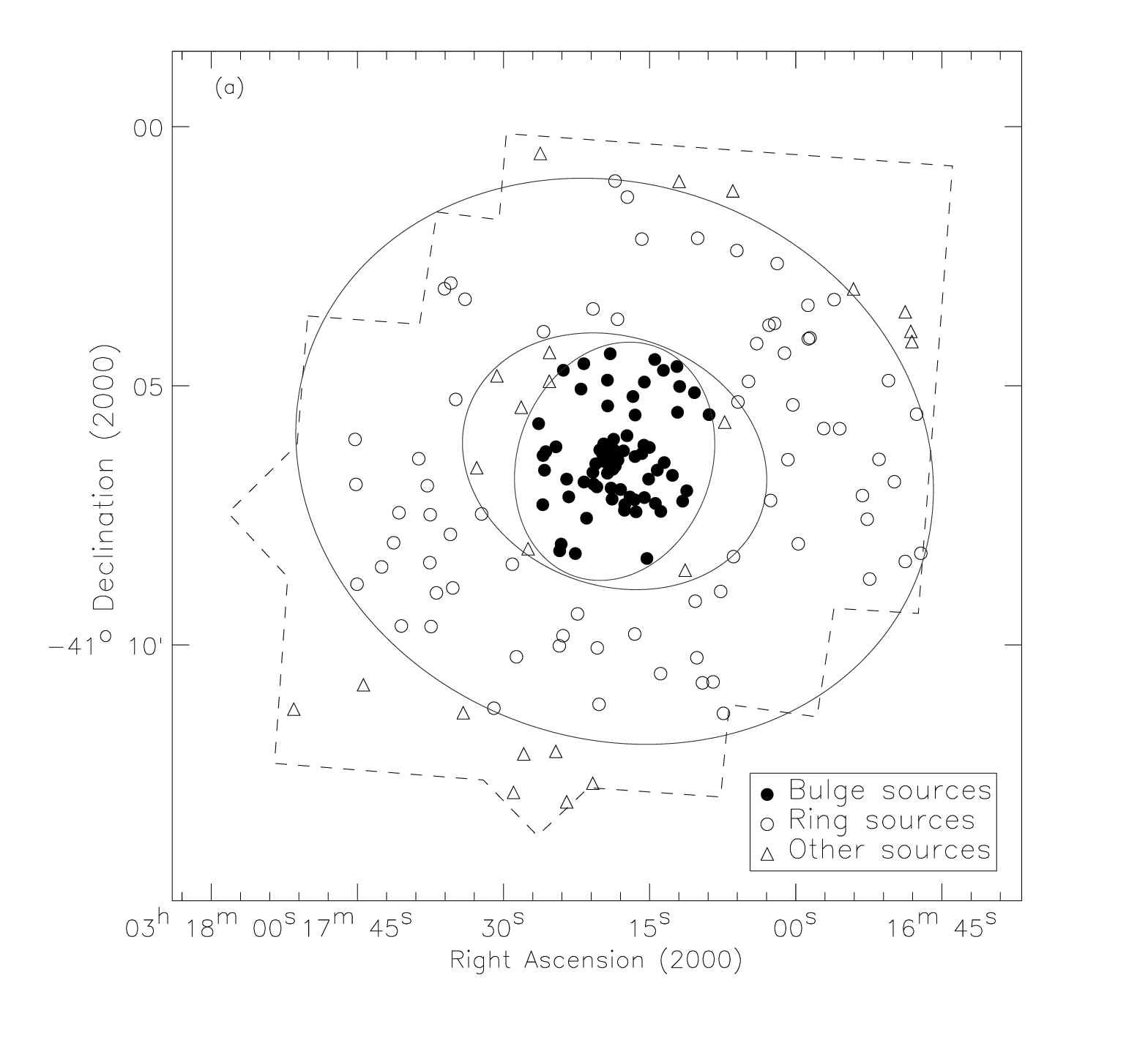}
\includegraphics[scale=0.5]{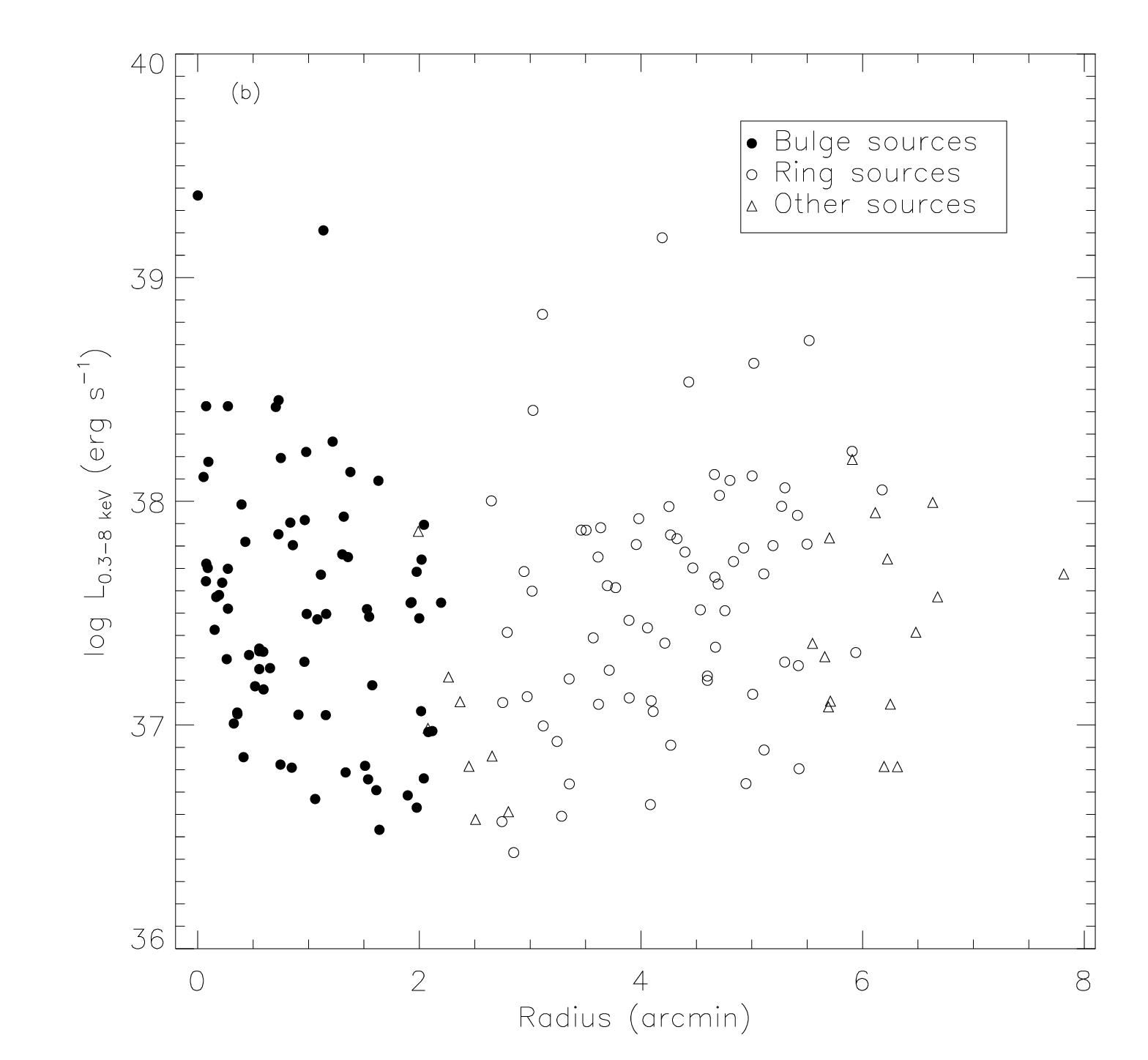}
}
\figcaption{
(a) Point-source spatial distributions for the 75 bulge sources (filled dots),
71 ring sources (open dots), and 23 other sources (triangles).
The dashed boundary lines shows the ACIS-S3
coverage of NGC 1291.
(b) Full-band luminosity vs. distance to the nucleus of the galaxy.
\label{posplot}}
\end{figure*}

The spatial distribution of the X-ray sources is 
displayed in Figure~\ref{posplot}a,
and their luminosity distribution is show in Figure~\ref{posplot}b.
The full-band luminosities range from $2.7\times10^{36}$ to $2.3\times10^{39}$~\lum.
We listed basic photometric properties for the 169 sources in Table 2,
with details of the columns given below.

\begin{enumerate}

\item
Column~1: the source number, listed in order of
increasing right ascension.

\item Columns~2 and 3: the right ascension and declination of
the \hbox{X-ray} source, respectively. 

\item Column~4: the off-axis angle of
the \hbox{X-ray} source in arcminutes. This is
the distance between the source position given in columns~2 and 3 and the
average aim point (see Section~\ref{datared}).

\item 
Column~5: the radial distance of 
the \hbox{X-ray} source to the nucleus of the galaxy in arcminutes. 

\item Columns~6--23: the source counts and the corresponding $1\sigma$
statistical errors or the upper limits on source counts
for the six bands, \hbox{0.3--8.0~keV}, \hbox{0.3--2.0~keV},
\hbox{0.5--2.0~keV},
\hbox{2.0--8.0~keV}, \hbox{0.3--1.0~keV},
and \hbox{1.0--2.0~keV}, respectively.
Upper limits are indicated as a ``$-1.0$'' in the error columns.

\item Column~24: the full-band effective exposure time determined from the
exposure map, which has been corrected for
vignetting and quantum-efficiency degradation.

\item Columns~25--30: the soft and hard X-ray colors and the 
corresponding $1\sigma$ statistical errors; see Section~\ref{xcolor} below for 
details.

\item Column~31: the photon index derived from spectral fitting, or set
to $\Gamma=1.7$ if the number of full-band counts is smaller than 30.

\item Column~32: observed full-band X-ray luminosity, in units
of \lum.

\item Column~33: the statistical significance of the full-band
flux variability; see Section~\ref{var} below for
details. It is set to ``$-1.0$'' if the flux variability is not available.

\item Column~34: region flag for the source. The flag is ``Bulge'' if
the source is located in the bulge, ``Ring'' if the source is in the ring,
or ``Other'' if the source is neither in the bulge nor in the ring.

\end{enumerate}

\section{POINT-SOURCE PROPERTIES}

\subsection{Radial Profile} \label{radprofilesec}

\begin{figure}
\centerline{
\includegraphics[scale=0.5]{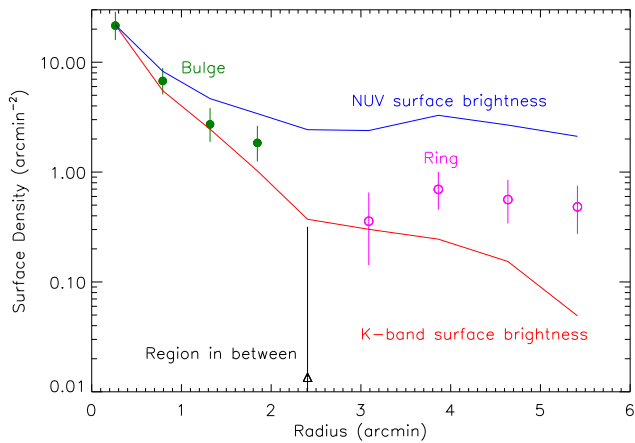}
}
\figcaption{
Radial profile of the X-ray source surface density.
The densities were computed in four elliptical annular regions in the bulge
(green filled circles),
four elliptical annular regions in the ring (magenta open circles), 
and one region in between
the bulge and the ring (black triangle). 
The $x$-axis shows the average distance of each region to the galactic nucleus.
The surface densities have been corrected for detection incompleteness
and background AGN contamination.
Only sources outside the central 10\arcsec-radius region 
and having $L_{\rm X}>10^{37}$~\lum\ are included in the density calculation.
The errors of the surface densities were calculated based on the Poisson
errors of the number of sources in each bin \citep{Gehrels1986}.
The red and blue curves display 
the surface brightness radial profiles (scaled to the highest X-ray 
source density)
for the {\it GALEX} NUV and 2MASS $K$-band data, respectively.
There appears to be an overdensity of X-ray sources in the  
ring region.
(A color version of this figure is available in the online journal.)
\label{radprofile}}
\end{figure}

We constructed a radial profile of the X-ray source surface density.
Given the non-circular shapes of the ring and bulge, we calculated
X-ray source densities in four elliptical annular regions in the bulge,
four elliptical annular regions in the ring, and one region in between
the bulge and the ring. The elliptical annular regions in the ring/bulge
have the same position angle and axis ratio as the ring/bulge 
and are evenly spaced in their distances to the galactic nucleus.
We corrected the number of X-ray sources in each region for the 
detection incompleteness and background AGN
contamination (see Section \ref{completeness} below; the incompleteness
correction was derived in the corresponding region). For this purpose,
we only included
sources that are outside the central 10\arcsec-radius region and have
full-band luminosity $L_{\rm X}>1\times10^{37}$~\lum. 
The corrected number of sources was 
then divided by the region area to obtain the surface density.
The X-ray source radial profile is presented in Figure~\ref{radprofile}.
There appears to be an overdensity of X-ray sources in the 
ring of the galaxy compared to the region in between the bulge and the ring.
We also included in this plot surface brightness radial profiles
for the {\it GALEX} NUV and 2MASS $K$-band data, calculated in the same 
regions as above. The surface brightness profiles are arbitrarily 
normalized so that the highest brightness has the same value as the highest
source density.
The NUV surface brightness is expected to trace the young 
stellar populations, while the $K$-band emission is sensitive to the old
stellar populations. 
In the ring region, the NUV and $K$-band surface brightness profiles differ 
significantly. The X-ray source radial profile in the ring 
does not continue with
the trend shown in the bulge and the region in between,
and it mainly follows 
the NUV surface brightness distribution, 
suggesting that the X-ray sources in the bulge and ring regions are 
associated with different stellar populations and the ring stellar population
is likely younger.

\subsection{Central Point Source and Ultraluminous X-Ray Sources} 
\label{nusource}
\begin{figure}
\centerline{
\includegraphics[scale=0.5]{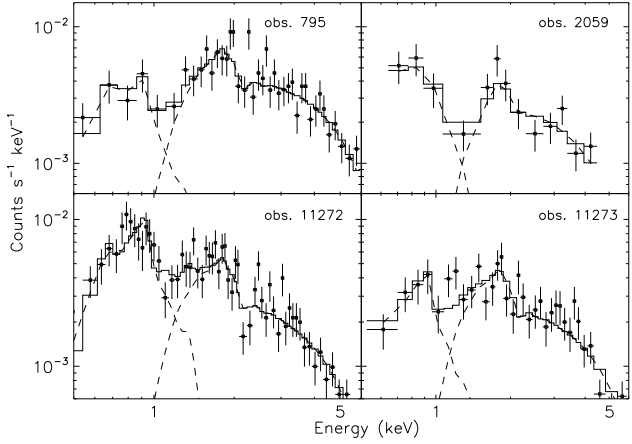}
}
\figcaption{
X-ray spectra of the nuclear source
overlaid with the best-fit models for the four individual observations.
The spectra were fit with an absorbed power law plus thermal plasma
model ({\sc wabs1*apec+wabs2*pow}).
The dashed curves show the individual model components.
For the thermal component, 
the temperature ($0.15$~keV)       
and column density ($7.6\times
10^{21}$~cm$^{-2}$) were
fixed at the values derived from the stacked spectrum.
See Section \ref{nusource} for more details and 
see Table~\ref{specfit} for the fit parameters.
\label{lda}}
\end{figure}

\begin{figure}
\centerline{
\includegraphics[scale=0.5]{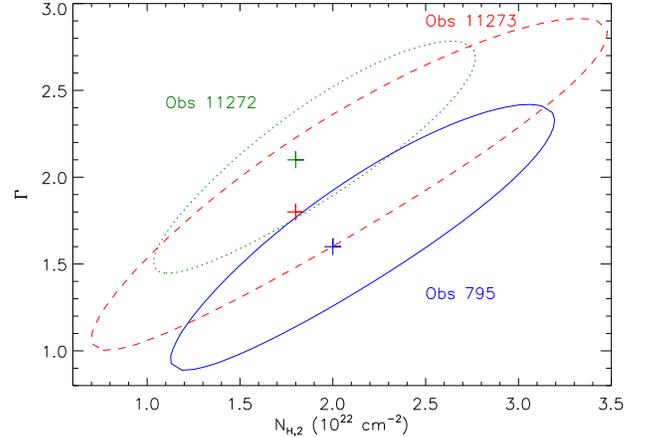}
}
\figcaption{
3$\sigma$ confidence contours for 
the photon index and absorption column density parameters of the nuclear
AGN in the three individual observations,
observations 795 (blue), 11272 (green), and 11273 (red).
The fit parameters in observation 2059 have large uncertainties
and is not shown in the plot for clarity.
The crosses show the best-fit 
$N_{\rm H,2}$ and $\Gamma$ values in observations 795, 11272, and 11273,
respectively. The larger confidence contour of observation 11273
than that of observation 795 despite of its longer exposure is due to 
the lower flux/luminosity of the nucleus during this observation.
See Section \ref{nusource} for more details.
(A color version of this figure is available in the online journal.)
\label{con}}
\end{figure}

The nuclear point source (XID 84) of NGC 1291 has a position of 
$\alpha_{\rm J2000.0}=
03^{\rm h}17^{\rm m}18.58^{\rm s}$, $\delta_{\rm J2000.0}=-41\degr06\arcmin28.7\arcsec$, and an observed full-band luminosity of $2.3\times10^{39}$~\lum.
It has been classified as a  
low-ionization nuclear emission-line region (LINER; e.g., \citealt{Smith2007}),
and \citet{Irwin2002} suggested that it is an obscured low-luminosity
active galactic nucleus (AGN) 
based on the X-ray spectral analysis of the first two \chandra\ observations.
We show the X-ray spectra of the source for the four individual observations
in Figure~\ref{lda}, which were extracted by AE with polygonal regions
that approximate the
$\approx89\%$ EEF PSFs in the 0.3--8.0 keV band. 
Local background spectra were also extracted by AE in the same annular
regions as those used in the photometry extraction.
There are clearly strong soft X-ray excesses in the spectra of
all the observations, which are typical among low-luminosity AGNs 
and are considered to originate from hot gas in the galactic center 
\citep[e.g.,][]{Ptak1999}.
We thus fit the spectra using
XSPEC with an 
absorbed power-law (AGN) plus thermal plasma (hot gas) model 
({\sc wabs1*apec+wabs2*pow}).
Note that there could be multiple
temperatures present in the gas, however, the limited signal-to-noise 
ratio (S/N) of the data cannot place strong constraints on a
multi-temperature plasma. 
The spectral slope and column density for the power-law component
were free to vary during the fitting.
For the thermal component, 
we set the abundance to be Solar,
and then we determined its 
temperature ($0.15_{-0.01}^{+0.04}$~keV) and absorption 
column density ($N_{\rm H,1}=7.6_{-0.7}^{+0.8}\times
10^{21}$~cm$^{-2}$)
by fitting the stacked spectrum of the 
four observations.
In the spectral fitting of the individual observations,
the temperature and column density 
parameters were fixed, and only the normalization was allowed to 
vary.\footnote{The extraction regions are different between observations,
corresponding to the $\approx89\%$ EEF PSFs of the nuclear point source 
in individual observations.
Therefore, the amount of diffused gas within the extraction region
varies.} 
The best-fit parameters are listed in Table~\ref{specfit};
also shown are the observed luminosities of the AGN component.
The 1$\sigma$ errors were derived by varying one interesting parameter.
The intrinsic 2.0--8.0 keV X-ray luminosity is
$\approx (2\textrm{--}3)\times10^{39}$~\lum\ after absorption and
aperture corrections, indicating
its low-luminosity
nature.
The nuclear AGN has an intrinsic power-law photon index of $\approx2$
and an absorption column density of 
$\approx2\times 10^{22}$~cm$^{-2}$, which are typical for an 
obscured AGN \citep[e.g.,][]{Turner1997}. 
The 3$\sigma$ confidence contours for the photon index
and column density parameters are shown in Figure~\ref{con};
there is a $\approx3\sigma$ spectral variation
between observations 795 and 11272. The X-ray luminosities of these
two observations also differ at the $\approx3\sigma$ significance level
(Table 1).
It has been noted that 
long-term variability in luminosity and spectral parameters
is common among obscured AGNs \citep[e.g.,][]{Risaliti2002}.

Besides the nuclear source, there are two more sources (XIDs 55 and 57) 
with observed full-band
luminosity in excess of $10^{39}$~\lum, in the regime of
ULXs. Their intrinsic full-band
luminosities after absorption corrections 
are $2.1\times10^{39}$~\lum\ and $2.0\times10^{39}$~\lum,
respectively.
One other source (XID 26),
after being corrected for intrinsic absorption 
($N_{H}=1.4\times 10^{21}$~cm$^{-2}$), has an intrinsic full-band
luminosity of $1.1\times10^{39}$~\lum, which should also be considered
as a ULX candidate. XIDs 26 and 55 are ring sources, 
whereas XID 57 is a bulge source.
The two ring ULX candidates have been reported in the ULX catalog of \citet{Swartz2004}
based on the data of the first observation (observation 795). 
The bulge ULX candidate is a transient source,
not detected in the first two observations. These ULX candidates 
were not detected 
in the study of \citet{Irwin2002}, which utilized only the first observation 
and focused on the bulge region. There is also a non-negligible possibility 
that these sources are 
background AGNs. The expected numbers of background AGNs 
with ULX luminosities if placed at the distance of 
NGC 1291 are $\approx0.05$ and $\approx0.3$
for the bulge and ring regions, respectively (see Section \ref{completeness}
below
for details of the AGN surface density).
The numbers of ULXs in NGC 1291 and the Cartwheel galaxy ($\approx3$ and 
$\approx20$) do not appear to be linearly correlated with their
SFRs (0.4~$M_{\sun}$~yr$^{-1}$ and 67.5~$M_{\sun}$~yr$^{-1}$).
Previous ULX surveys have found that the 
numbers of ULXs at low SFRs ($\le0.4$~$M_{\sun}$~yr$^{-1}$)
are significantly higher than those extrapolated linearly from 
the high SFR end, suggestive of an additional population of 
ULXs associated with LMXBs \citep[e.g.,][]{Liu2006}.

\subsection{Source Variability} \label{var}

X-ray flux/spectral variability is a common feature among point sources in
galaxies, which is generally attributed to the change of physical
properties of the accretion disks
\citep[e.g.,][]{Remillard2006,Done2007,Brassington2010,Fabbiano2010}.
The four \chandra\ observations of NGC 1291 span $\approx10$ years,
allowing us to study the long-term flux variability of the point sources.

\begin{figure}
\centerline{
\includegraphics[scale=0.5]{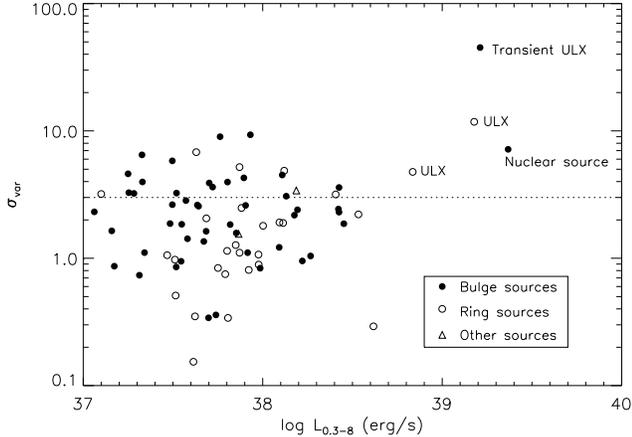}}
\figcaption{
Significance of flux variability vs. observed full-band luminosity for the 
50 bulge sources (filled dots),
28 ring sources (open dots), and 3 other sources (triangles). 
Dotted line indicates the $>3\sigma$ significance level of the 
flux variability.
Note 
that we calculated flux variability for only sources detected 
in at least one individual observation; there are insufficient data
for variability study of the remaining sources. We studied 
spectral variability of the few ULX candidates, 
which show no significant variability in spectral
parameters (column density and power-law index).
\label{fluxvar}}
\end{figure}

For each source in each observation, we calculated its aperture-corrected
full-band count rate ($CR$) based on the photometry provided by AE;
an upper limit on the count rate was calculated if the source was not
detected in the observation (see Section \ref{catalog}).
For a source that was detected in at least one individual observation and
was covered by at least two observations, we
computed the maximum statistical significance of its flux variation between
any two observations, defined as 
\citep[e.g.,][]{Brassington2009,Sell2011}:
\begin{equation}
\sigma_{var}={\rm max}_{i,j}\frac{|CR_i-CR_j|}{\sqrt{\sigma_{CR_i}^2+\sigma_{CR_j}^2}}~,
\end{equation}
where the subscripts $i$ and $j$ run over different observations, and $\sigma_{CR_i}$ and $\sigma_{CR_j}$ are the 1$\sigma$ errors of the 
count rates. 
We consider a source to be variable if its $\sigma_{var}$
parameter is greater than three (i.e., $>3\sigma$ variation).
The flux variability for the point sources in NGC 1291 are 
shown in Figure~\ref{fluxvar}. In the bulge and ring regions,
$38_{-9}^{+11}\%$ and $25_{-9}^{+13}\%$ of 
the sources are variable, respectively; the errors are 
the 1$\sigma$ Poisson uncertainties on the numbers of sources 
that are considered
variable. We note that the central point source and the three ULX candidates 
are
variable.
There is no significant difference between the fraction of 
variable sources in the bulge and ring regions given the data available.
\citet{Kilgard2005} reported that 29\%
of the sources show long-term variability based on the first two
\chandra\ observations, consistent with our findings here.
The fraction of variable sources in NGC 1291 is in general agreement
with those in some typical early-type galaxies, e.g., NGC 3379 and NGC 4278
\citep{Brassington2008,Brassington2009}.

For the luminous nuclear point source and the three ULX candidates, we also 
examined their 
spectral variability by fitting the spectra in individual observations.
The spectral variability of the nuclear source is discussed in 
Section~\ref{nusource}.
For the ULX 
candidates,
their spectra were well fit 
by an absorbed power-law model. 
Although all these
ULX candidates are variable in flux, none of them
shows significant ($>3\sigma$) variability in its spectral
parameters. The two non-transient ULX candidates (XIDS 26 and 55) still
have ULX luminosities in individual observations.

\subsection{X-Ray Colors} \label{xcolor}

\begin{figure}
\centerline{
\includegraphics[scale=0.5]{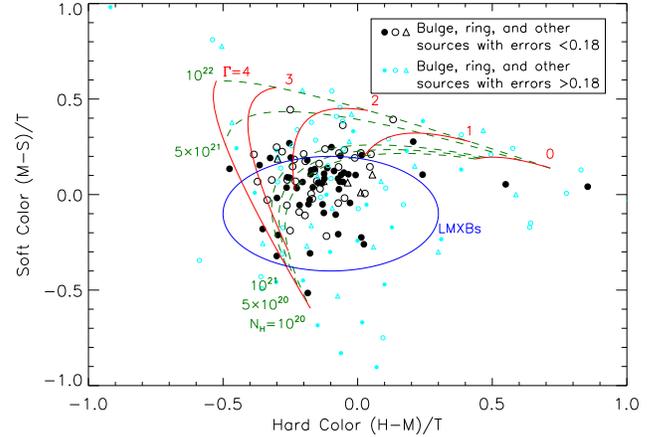}} 
\figcaption{
X-ray color-color plot for the point sources. $S$, $M$, $H$, and $T$ are the
count rates in the 0.3--1.0 keV, 1.0--2.0 keV, 2.0--8.0 keV, 
and 0.3--8.0 keV bands, respectively.
Black data points represent sources with relatively small color errors, while
cyan data points represent sources with large errors and are less significant.
The green and red tracks show the expected X-ray colors for absorbed power-law
spectra with different power-law indices and column densities.
The blue ellipse indicates the area that is likely dominated by LMXBs  
\citep{Prestwich2003}.
A significant fraction
of the X-ray sources with small color errors are located in 
the region expected to be dominated by LMXBs.
(A color version of this figure is available in the online journal.)
\label{hardplot}}
\end{figure}

It is difficult to classify individual X-ray sources into different
XRB types (e.g., LMXBs and HMXBs) given the limited X-ray
data alone \citep[e.g.,][]{Fabbiano2006}.
However, we can use X-ray colors to separate 
the points sources into groups that are likely
dominated by certain source types \citep[e.g.,][]{Colbert2004,Prestwich2003,Prestwich2009}.
We constructed an X-ray color-color plot following \citet{Prestwich2003}.
The X-ray colors are defined as
$HC=(H-M)/T$ (hard color) and $SC=(M-S)/T$ (soft color), 
where $S$, $M$, $H$, and $T$ are the 
count rates in the 0.3--1.0 keV, 1.0--2.0 keV, 2.0--8.0 keV, and 
0.3--8.0 keV bands, respectively.
A Bayesian approach was adopted to calculate the colors and the 
associated errors following the method developed by \citet{Park2006}.
This approach provides a rigorous statistical treatment of the Poisson
nature of the detected photons as well as the non-Gaussian nature of 
the error propagation.
The color-color plot is shown in Figure~\ref{hardplot}.
We consider a source to be significant in the plot (a black data point) 
if its average error of the soft and hard colors is smaller than
the $3\sigma$-clipped mean (0.18) of the errors for all the sources.
The other sources (cyan data points) have relatively large color errors and 
are not included in the following analysis.
A significant fraction (75\% of the bulge sources and 65\% of the ring sources) 
of the X-ray sources are located in the region that is 
likely dominated by LMXBs \citep{Prestwich2003}. 
The dominance of LMXBs is also suggested by the low SFR of NGC 1291.
The SFR of NGC 1291 is 
$\approx20\%$ of that of the Milky Way \citep[e.g.,][]{Chomiuk2011},
and the Milky Way only hosts a few HMXBs with luminosities above our detection 
completeness limit ($\approx10^{37}$~\lum; see Section 5.1) given its 
population studies \citep[e.g.,][]{Grimm2002,Voss2010}. 
Therefore, HMXBs should have 
a small contribution to the observed XRB population in NGC 1291.

The bulge and ring sources appear to have different distributions in
the X-ray color-color space. 
A two-dimensional Kolmogorov-Smirnov test\footnote{We used
IDL routine {\sc ks2d} for this test; 
see http://www.astro.washington.edu/users/yoachim/idl/py\_idl.html.}
indicates that the probability of the two distributions 
being drawn from the same parent population is 0.08. 
The $3\sigma$-clipped mean
soft and hard colors for the bulge population are
$0.04\pm0.02$ and $-0.14\pm0.02$, and the mean values are
$0.11\pm0.02$ and $-0.15\pm0.02$ for the ring population.
On average, the ring population appears to have a 
significantly harder soft X-ray color 
than the bulge population. 
The color-color plot is overlaid with a grid showing the 
expected locations of sources with 
different power-law spectra and absorption column densities, derived
using the CXC's Portable, Interactive, Multi-Mission Simulator
(PIMMS).
A generally harder soft color suggests that the ring sources are more
obscured (larger $N_{\rm H}$) than the bulge sources, as would be expected
in regions with ongoing/recent star formation.

\section{X-RAY LUMINOSITY FUNCTIONS} \label{xlf}
We construct full-band 
XLFs for the bulge and ring regions
using the merged \chandra\ observation. Although $\approx25$--40\% of the 
point sources are variable between the individual observations, the XLFs
derived from the merged observation should be stable against 
source variability based on previous studies of the XLFs for various 
systems \citep[e.g.,][]{Grimm2005,Zezas2007,Sell2011}.

\subsection{Incompleteness Corrections and Background AGN Subtraction} \label{completeness}

To derive the XLFs correctly, it is essential to make proper corrections
for the detection incompleteness and background AGN contamination.
We performed simulations following \citet{Kim2004b} to correct for the
effects of incompleteness,
including source detection limit and source
confusion. Briefly, in each simulation we added a mock 
source at a random location
of the event file of every observation using the MARX ray-tracing
simulator,\footnote{http://space.mit.edu/CXC/MARX/index.html} 
created a merged event file and a merged full-band image,
and then ran
{\sc wavdetect} to test if this additional source is detectable.
The input 
X-ray luminosity of the source was randomly drawn from a power-law XLF with
$\beta=1$ in a cumulative form, $N(>L_{\rm X})=kL_{\rm X}^{-\beta}$.
We assumed a power-law spectrum for the source with $\Gamma=1.7$.
The position of the source was randomly determined following the $r^{1/4}$ law
\citep{deVaucouleurs1948}. We note that the adopted luminosity and position
distributions here do not affect the incompleteness corrections significantly,
as we only aimed to derive the positional-dependent
detection fractions at a given luminosity.
We performed 60,000 simulations in total, and the incompleteness corrections
were calculated as a function of X-ray luminosity for the bulge 
and ring regions,
respectively, utilizing all the simulated sources in each region. 
We also excluded the central 10\arcsec-radius area from the bulge region
when computing XLFs as the incompleteness corrections are not reliable in
this crowded region.
The resulting 90\% (50\%) completeness limits are $\approx1.5\times10^{37}$
($\approx7.1\times10^{36}$) and
$\approx2.2\times10^{37}$ ($\approx8.5\times10^{36}$)~\lum\ for the bulge and ring populations, respectively. 
The completeness limits in the bulge are lower because of 
the longer effective exposure time in the bulge and
the smaller 
background within the extraction regions of bulge sources.
The XLFs were only calculated down to the
$50\%$
completeness limits, below which the Eddington bias \citep{Eddington1913}
could have affected the measured X-ray luminosities.
There are 64 and 61 sources with luminosities above these completeness limits
in the bulge and ring regions, respectively; $\approx98\%$ of the sources
have more than 10 full-band counts and thus the Eddington bias should have
minimal effect \citep[e.g.,][]{Bauer2004}.  
Compared to the XLFs presented in \citet{Irwin2002} and \citet{Kilgard2005}, 
which were based on the first two \chandra\ observations,
we are probing the XLFs about twice as deep.

To correct for background AGN contamination, we did not 
attempt to classify AGNs from the detected
sources. Instead, we calculated the expected AGN luminosity function, which 
was then deducted from the observed XLFs.
We computed the AGNs flux
distribution using the \citet{Gilli2007} AGN PS model, and we
normalized
the AGN surface density to the observed value in the $\approx4$~Ms \chandra\ Deep
Field-South, which is $\approx10\,000$ deg$^{-2}$ at a flux limit of 
$\approx3.2\times10^{-17}$~\flux\ \citep{Xue2011}. 
A 20\% uncertainty was included for this estimation
that was attributed to cosmic variance \citep[e.g.,][]{Luo2008}.
There are $4.2\pm0.8$ and $21.5\pm4.3$ AGNs expected in the bulge and ring
regions, respectively. The ring XRB population is heavily contaminated 
($\approx35\%$)
by background AGNs. 
To derive the differential XLFs of the point sources in NGC 1291, we calculated the expected
number of background AGNs in a given luminosity bin 
and then subtracted it from the incompleteness-corrected number of 
X-ray sources in this bin. The cumulative XLFs were constructed by 
integrating the differential XLFs.

\begin{figure*}
\centerline{
\includegraphics[scale=0.5]{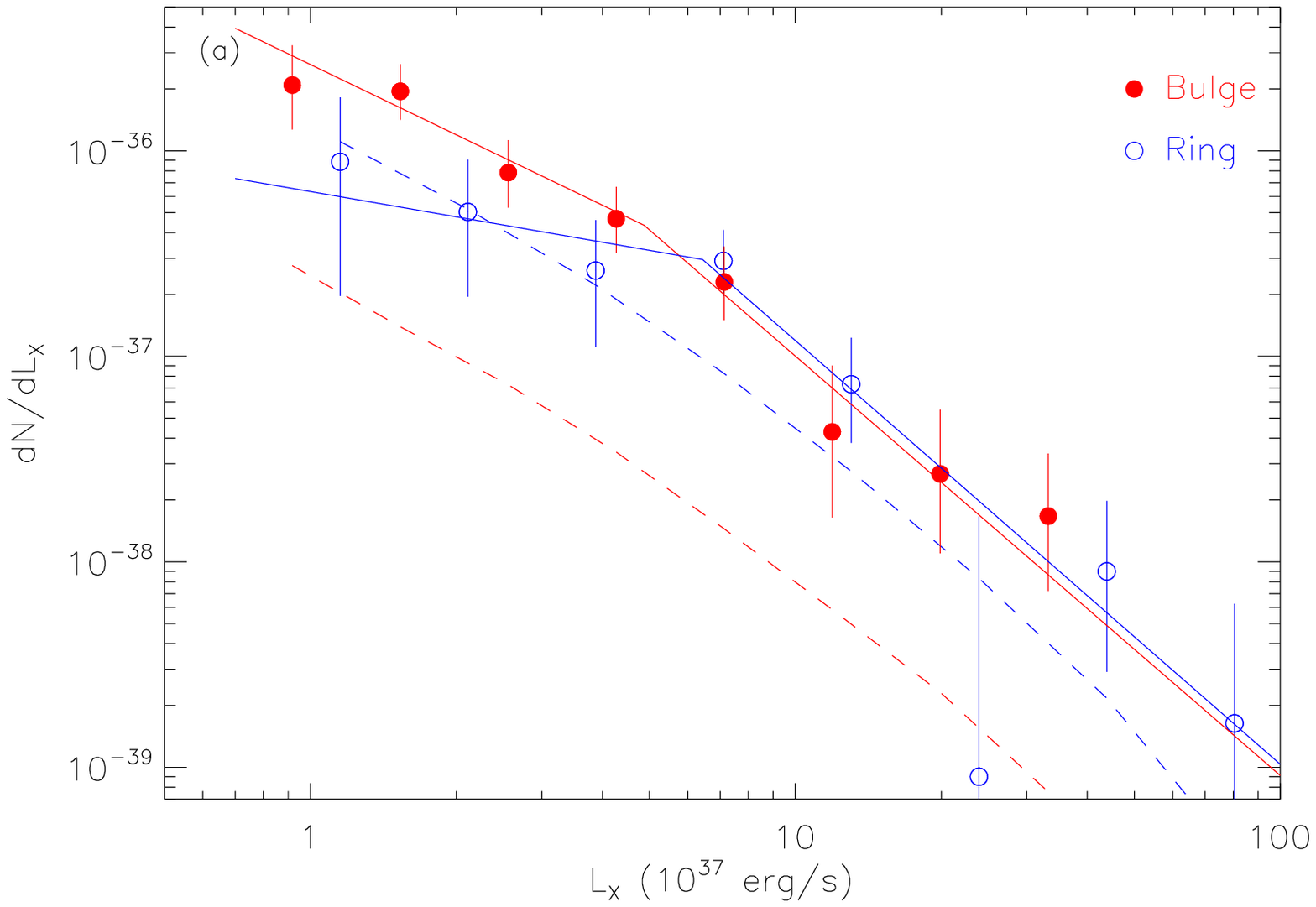}
\includegraphics[scale=0.5]{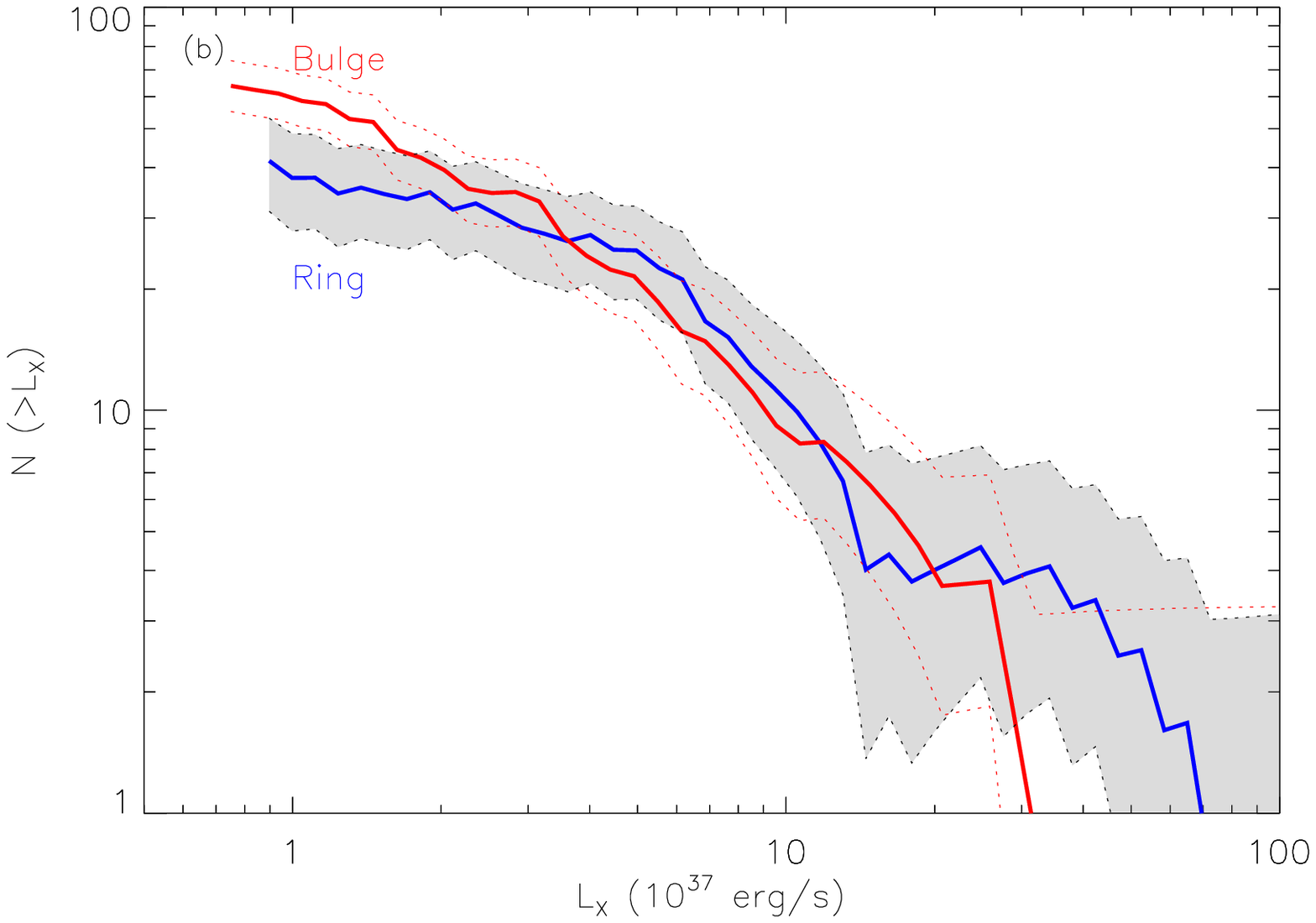}}
\figcaption{Differential [panel (a)] and cumulative [panel (b)] XLFs
for the bulge (red filled dots or curve)
and ring (blue open dots or curve) populations. 
The XLFs have been corrected for detection incompleteness
and AGN contamination.
In panel (a), the dashed curves represent 
the expected XLFs of background AGNs for the corresponding regions.
The ring XRB population is heavily contaminated
by background AGNs, resulting in large error bars for the XLF data points.
The solid curves are the best broken power-law fits to the corresponding
XLFs.
In panel (b), the red dotted curves represent the 1$\sigma$ errors
for the ring XLF, and the grey region represents the 1$\sigma$ errors
for the bulge XLF. The cumulative XLFs do not appear to
monotonically decrease with luminosity at a few places due to
the uncertainties introduced by the background subtraction.
The XLFs were computed down to the $\approx50\%$
completeness limits.  
(A color version of this figure is available in the online journal.)
\label{xlfplot}}
\end{figure*}

The incompleteness-corrected and background subtracted XLFs for the bulge
and ring populations are displayed in Figure~\ref{xlfplot}.
The 1$\sigma$ errors were determined by combining the Poisson
errors of the number
of sources in each luminosity bin and the 20\% uncertainty of the AGN
surface density.

\subsection{Parameterization of the X-Ray Luminosity Functions} \label{xlffit}

The shapes of XLFs for different XRB
populations (mainly LMXBs and HMXBs) have been studied
extensively in the literature
(see e.g., \citealt{Fabbiano2006} for a review). For LMXBs,
the XLF in the bright end ($L_{\rm X}>$ a few $10^{37}$~\lum)
can usually be parameterized
with broken power laws. The break luminosity is
$\approx(2\textrm{--}5)\times 10^{38}$~\lum, which may be related
to the Eddington luminosity of neutron star binaries
\citep[e.g.,][]{Sarazin2000,Finoguenov2002,Kim2004b}.
Below the break luminosity,
the XLFs have differential power-law slopes of 1.8--2.2
\citep[e.g.,][]{Kim2004b}; above the break, the XLFs cut off sharply,
with differential slopes of $\ga2.8$.
Note that the XLFs for young ($<5$ Gyr) LMXBs samples may not
present this high-luminosity break \citep{Kim2010}.
In the faint end
($L_{\rm X}<$ a few $10^{37}$~\lum), there are limited data available
for the XLFs of LMXBs.
The XLFs of nearby early-type galaxies and the bulges of nearby spirals
(including the Milky Way) suggest that there is a second break
at $\approx(2\textrm{--}5)\times 10^{37}$~\lum, below which the XLFs
appear flatter with differential slopes of $\approx 0.5$--1
\citep[e.g.,][]{Gilfanov2004,Voss2007,Kim2009,Zhang2011}.\footnote{
It has been suggested that LMXBs in different environments
(e.g., in the field or globular clusters)
probably exhibit notable difference in their XLFs \citep[e.g.,][]{Kim2009}.
Such detailed discussion is beyond the scope of the current paper.}
The origin of this low-luminosity break is not clear,
which could be related to magnetic braking during binary stellar evolution
\citep[e.g.,][]{Postnov2005,Fragos2008} or
different types of donor stars in high- and
low-luminosity LMXBs \citep{Revnivtsev2011}.
For HMXBs, the XLFs can be described with a simple power law
with a differential slope of $\approx1.6$ over a large luminosity
range ($\approx5\times10^{36}\textrm{--}10^{40}$~\lum; \citealt{Grimm2003}).

We fit the XLFs of NGC 1291 to parameterize their shapes. A single
power law of the form
\begin{equation}
\frac{dN}{dL_{\rm X}}=A1\left(\frac{L_{\rm X}}{10^{37}}\right)^{-\alpha}
\end{equation}
and a broken power law of the form
\begin{equation}
\frac{dN}{dL_{\rm X}}=
\begin{array}{ll}
A2\left(\frac{L_{\rm X}}{L_{\rm b}}\right)^{-\alpha_1} & \text{if~} L_{\rm X}\le L_{\rm b} \\
A2\left(\frac{L_{\rm X}}{L_{\rm b}}\right)^{-\alpha_2} & \text{if~} L_{\rm X}> L_{\rm b}
\end{array}
\end{equation}
were used, where $A1$ and $A2$ are the numbers of sources at the reference
luminosites.
We used the {\sc sherpa} spectral fitting tool in the CIAO package
to fit the XLFs in their differential forms.
It has been shown in previous studies that 
this is an efficient method to derive the characteristics of XLFs and it can yield
consistent results with a maximum likelihood approach \citep[e.g.,][]{Zezas2007,Kim2009,Sell2011}.
The XLFs were read into {\sc sherpa} as typical X-ray spectra with the number
counts binned in small
luminosity bins, $\delta\log(L_{\rm X})=0.05$.
Given the small numbers of sources in the bulge and ring regions, we used the 
Cash statistic in {\sc sherpa} for the model fits. 
The incompleteness corrections were supplied as an ancillary response file 
(ARF), and the AGN contribution was supplied as a table model in
addition to the intrinsic power-law model. 
The best-fit parameters with 1$\sigma$ errors 
are reported in Table~\ref{tbl-fit}, and
the 
best-fit broken power-law models are shown in Figure~\ref{xlfplot}.
For the single-power model, both XLFs have a power slope
of $\approx1.5$. For the broken power-law model, both XLFs show
a break luminosity of $\approx5\times 10^{37}$~\lum. Above the 
break luminosity, the power slope is $\approx2$; below that,
the bulge XLF has a slope of $\approx1$ while the ring XLF appears flatter,
with a slope of $\approx0.4$. Note that the errors for the broken power-law
parameters are relatively large.

The best-fit parameters were derived using the
Cash
statistic, which does not provide a goodness-of-fit parameter analytically.
In order to compare the quality of the power-law and broken power-law fits,
we adopted the likelihood ratio test used in \citet{Zezas2007}.
For each differential XLF (ring/bulge), we simulated 10,000 mock 
XLFs from the best-fit power-law model, 
using the same luminosity bins and adding Poisson fluctuations to the
source counts. The
detection incompleteness and background AGN contamination were also applied
to the mock XLFs. We then fit the simulated XLFs with both the
power-law and broken power-law models. The ratio of the Cash statistic
values of the best power-law and broken power-law fits for the 
simulated XLFs were calculated and compared to the ratio for the 
observed XLF; the rate of obtaining a higher ratio in the simulations 
represents the probability of getting an improved broken power-law fit
by chance. For the bulge XLF, we found that  
the broken power-law model provides a statistically improved
fit at the 89\% confidence level; for the ring XLF 
the broken power-law model provides an improved fit at the 
94\% confidence level.
Therefore, the broken power-law model appears to provide 
better fits
to the data for both the bulge and ring populations.
The broken power-law shapes of the XLFs suggest that both 
the bulge and ring XRB populations  
are dominated by LMXBs. We are probing the low-luminosity break at
$\approx5\times 10^{37}$~\lum, with a slope of $\approx2$ at the brighter
end and a flatter shape at the fainter end; there are too few luminous
XRBs in NGC 1291 to constrain the high-luminosity break at 
$\approx(2\textrm{--}5)\times 10^{38}$~\lum. This result is consistent with 
that inferred from the 
X-ray colors (see Section~\ref{xcolor}).
Given the arguments above, 
we prefer the broken power-law model for both the 
bulge and ring XLFs, and the following discussions are based on the
broken power-law parameters.

\section{THEORETICAL MODELING OF THE X-RAY BINARY POPULATIONS} \label{modeling}

PS modeling has been proven a valuable tool 
in order to study the physical properties of the XRB populations 
observed with \chandra\ in nearby galaxies. \citet{Belczynski2004} 
constructed the first synthetic XRB populations for direct comparison 
with the \chandra\ observed XLF of NGC 1569, a star forming dwarf 
irregular galaxy. \citet{Linden2009,Linden2010} 
developed models for the HMXB and Be XRBs of the SMC, 
studying the XLF and the spatial distribution of this population
and investigating the effect of electron-capture supernovae of 
massive ONeMg stellar cores. 
\citet{Fragos2008, Fragos2009} performed extensive PS simulations
for the modeling of the two old elliptical galaxies NGC 3379 and NGC 4278. 
They studied the relative contribution to the observed XLF from 
sub-populations of LMXBs with different donor and accretor types, 
and the effects of the transient behavior of LMXBs. 
All these studies have been pivotal in interpreting \chandra\ observations 
of nearby galaxies.

In this paper, we utilized a \emph{new} grid of 
PS models to interpret the XLF characteristics of
NGC 1291. 
In these models, we focused on LMXBs and HMXBs 
formed in the galactic field as products of the evolution of isolated 
primordial binaries (i.e. we did not take into account any dynamical effects). 
We performed our simulations with {\tt StarTrack} 
\citep{Belczynski2002,Belczynski2008}, an advanced PS code that has been 
tested and calibrated using detailed mass transfer star calculations and 
observations of binary populations, and incorporates all the important 
physical processes of binary evolution.

In the development of our models we incorporate our current knowledge about 
the characteristics of the stellar population in NGC 1291. This galaxy 
has a stellar mass of $\approx 1.5 \times 10^{11}$~$M_{\sun}$ and a 
low current SFR of $\approx0.4$~$M_{\sun}$~yr$^{-1}$
\citep{Kennicutt2003}.
We adopted the \citet{Noll2009} model for the star-formation history
of NGC 1291 including two exponential starburst events (see Section 1).
The metallicity of both populations is assumed to be solar.
Due to the similar
shapes of the bulge and ring cumulative XLFs and their relatively
large uncertainties,
it is not plausible to compare models to the individual XLFs.
We also note that the best fit results by \citet{Noll2009} about the
star-formation history of NGC 1291 have no spatial information.
Hence, our comparison between PS models and the observed XRB population
is limited to the combined (bulge plus ring) XLF of the galaxy.

We created a grid of 48 PS models varying parameters that are known 
from previous studies \citep{Fragos2008} to affect the shape and 
normalization of the resulting synthetic XLF. Namely, we varied 
the common envelope (CE) efficiency, the initial mass function, 
stellar wind strength, and finally we allowed for various outcomes 
of the CE phase, in particular the possible CE 
inspirals with Hertzsprung gap donors that terminate binary 
evolution barring the subsequent XRB formation \citep{Belczynski2007}. 
The latter parameter, which is known to affect
the evolution of massive binaries which are the progenitors 
of double compact objects \citep{Belczynski2007}, 
is for the first time explored in the context of XRB populations.
The various parameter values used are shown in table \ref{model_param}. 
We note that in our calculations we combined $\alpha_{\rm CE}$ 
and $\lambda$ into one CE parameter, where $\lambda$ is a measure of 
the central concentration of the donor and the envelope binding energy. 
In the rest of the text, whenever we mention the CE efficiency 
$\alpha_{\rm CE}$, we refer in practice to the product 
$\alpha_{\rm CE} \times \lambda$, effectively treating 
$\lambda$ as a free parameter \citep[see][for details]{Belczynski2008}.

The synthetic XLFs were constructed based on the method 
described in \citet{Fragos2008}, while transient systems were
modeled according to \citet{Fragos2009}. Furthermore, we 
calculated the errors in the synthetic XLF due to Poisson 
low number statistics (high luminosity end) 
and uncertainties in the bolometric correction 
factors (low luminosity end) (T.~Fragos et al. 2012, in prep.; 
P.~Tzanavaris et al. 2012, in prep.). 
In order to assess which are the preferred values 
of the different model parameters, we calculated the likelihood of 
the observed XLF given a model. 
Our statistical analysis takes into account not 
only the shape of the XLF \citep[as in][]{Fragos2008}
but the absolute normalization too.
Our maximum likelihood model has 
a low CE efficiency ($\alpha_{\rm CE}\approx0.1$) and the ``standard'' 
stellar wind prescription \citep[][and references therein]{Belczynski2010}. 
We also found that there is an apparent degeneracy between the slope 
of IMF and whether or not we restrict the outcome of a CE event 
to a merger when the donor star is in the Hertzsprung gap. 
A flatter IMF results in a larger black hole XRB (BH-XRB) population, 
while on the other hand, assuming that a CE with a Hertzsprung gap 
donor leads always to a merger, which prevents 
many of the binaries from evolving into a BH-XRB. 
Hence, a BH-XRB population consistent with observations could
either result from a steeper IMF (slope of -2.7) and CE prescription where 
we allow for all possible outcomes, even with a Hertzsprung gap donor, 
or it could result from a flatter IMF (slope of -2.35) and a 
CE model where a 
CE event with a Hertzsprung gap donor would always lead to a merger. 
In fact, the former case corresponds to our maximum likelihood model, 
while the latter to the model with the second highest likelihood.

\begin{figure*}
\centerline{
\includegraphics[scale=0.5]{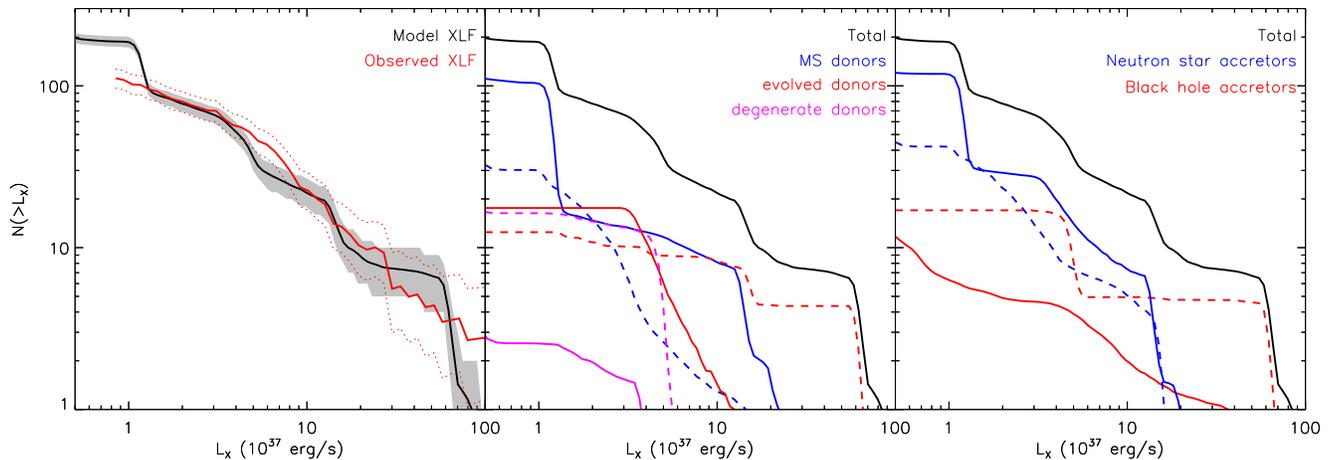}}
\figcaption{
\label{model_xlf} Left panel: Synthetic XLF (black solid curve) 
for the maximum likelihood model. The gray shaded area corresponds 
to the $1-\sigma$ error of the XLF due to Poisson statistics 
and uncertainties in bolometric correction. For comparison the 
observed combined (corresponding to both the bulge and ring regions) 
XLF (red solid curve) and the associated $1-\sigma$ error 
(red dotted curves) are drawn. Center panel: 
Contribution of different sub-populations 
by separating the XRBs into groups of systems 
with different donor stellar types. 
Right panel: Contribution of different sub-populations 
by separating the XRBs into groups with different accretor types.
In the
center and left panel, solid colored lines denote persistent XRB
sub-populations, while dashed colored lines denote transient ones.
(A color version of this figure is available in the online journal.)
}
\end{figure*}

Figure \ref{model_xlf} shows the synthetic XLF from our maximum likelihood 
model in comparison with the observed XLF, as well as the different 
types of XRBs contributing to the XLF at different ranges 
of X-ray luminosity. 
There are two artificial ``jumps'' in the model XLF at
$\approx10^{37}$~\lum\ and $\approx8\times10^{38}$~\lum, 
instead of the observed smoother power-law shape.
In our models, we identify transient
and persistent sources by comparing the calculated 
mass-transfer rate of the binary to a critical
mass-transfer rate, below which the thermal instability develops. 
Furthermore, we strictly limit
the accretion rate to the Eddington limit. 
These limiting mass-transfer rates in our modeling are
responsible for the apparent jumps in the synthetic XLF. 
However, in nature there is no sharp
transition between thermally stable and unstable disks, 
nor a precise limit in the highest
accretion rate possible. Accretion onto a compact object 
is a non-linear and much more complex
process, which in reality can results in a smoother 
luminosity distribution with no sharp
transitions. Finally, by assuming a solar 
metallicity for the stellar population of NGC 1291 in
order to be consistent with \citet{Noll2009} 
and in the absence of any stellar metallicity
measurements, we impose in practice a maximum BH 
mass of $\approx15~M_{\sun}$. If instead
part of the stellar population had a lower 
metallicity (e.g. 30\% solar) then the maximum BH mass
would increase to $30~M_{\sun}$ or more 
\citep{Belczynski2010}, which could smooth out the
jump at the very luminous end of the XLF.

We found that the BH-XRBs that populate the high-end of the XLF (above 
$\approx2\times 10^{38}$~\lum, Eddington limit for a hydrogen accreting
neutron star) are transient systems with evolved companion stars, while at lower
luminosities the XRB population consists of a mixture of transient and
persistent XRBs with a neutron star accretor and a giant or a main sequence
(MS) donor star.
These findings 
and the general shape of the synthetic XLF are in agreement with 
the PS study by \citet{Fragos2008} of the two old elliptical 
galaxies NGC 3379 and NGC 4278. This leads us to the conclusion 
that an old LMXB population is dominating the observed XLF. 
However, there is an evident overabundance of 
super-Eddington ($L_{\rm X} \ga 2\times 10^{38}$~\lum, the Eddington limit 
for a hydrogen accreting $1.4~M_{\sun}$ neutron star) XRBs, compared to 
the \citet{Fragos2008} models, which can be attributed to the 
enrichment of the younger XRB population with 
black hole accretors. 
This younger population is likely originated from the ring star-forming region. 
Our finding that the younger stellar population of the ring 
region enhances the high X-ray luminosity,
super-Eddington BH-XRB population, is consistent with the recent 
observational study by
\citet{Kim2010}. \citet{Kim2010} compared the combined 
XLF of LMXBs detected in \chandra\
observations of post-merger elliptical galaxies with that 
of typical old elliptical galaxies and
found that elliptical galaxies with signs of recent star formation host a larger fraction of
luminous LMXBs ($L_{\rm X} \ga 5\times 10^{38}$~\lum) than old elliptical
galaxies.

\section{DISCUSSION}

\subsection{The Bulge LMXB Population}

Galactic bulges host relatively homogeneous samples of
old stellar populations, and thus they are ideal targets for
the research of LMXB XLFs.
PS modeling of the XLFs
has been used to
constrain the formation, evolution, and physical properties
of the LMXB population
\citep[e.g.,][]{Fragos2008,Fragos2009}.
Recently deep \chandra\ exposures of nearby elliptical galaxies and spiral
bulges have pushed the XLFs down to limiting luminosities of around a few
$10^{36}$~\lum; some best studied examples include
Centaurus A \citep{Voss2006}, M31 \citep{Voss2007}, M81 \citep{Sell2011},
and samples of galactic bulges in \citet{Gilfanov2004} and \citet{Kim2009}.
The derived XLFs are remarkably similar,
with a low-luminosity break at $\approx5\times 10^{37}$~\lum\
and a high-luminosity break or cut-off at $\approx5\times 10^{38}$~\lum.
The bulge of NGC 1291 is dominated by LMXBs based on the XLF shape and the
X-ray colors, though
relatively 
weak star formation could still exist given the observed 
filaments of H$\alpha$ gas (See Section 1).
The bulge XLF of NGC 1291 reaches 
a similar threshold luminosity of $7\times10^{36}$~\lum, 
and its shape is statistically consistent
with previous findings. We obtain a low-luminosity break at 
$\approx5\times 10^{37}$~\lum; a high-luminosity break is not constrained
but is highly probable as there is only one bulge source with
$L_{X}>5\times 10^{38}$~\lum\ (one of the ULX candidates).
The power-law slope above the low-luminosity
break is $\approx2$, and is $\approx1$ below the break, typical for 
LMXB XLFs. 
Despite of the complex morphology of the galaxy, the bulge of NGC 1291
still hosts a population of LMXBs that is very similar to those in 
typical elliptical galaxies and spiral bulges 
(in terms of their XLFs)
down to the limiting 
luminosity, suggesting that 
all LMXB populations are compatible.

\subsection{Comparison of the Bulge and Ring Populations}

X-ray observations of the XRBs 
can provide valuable constraints on the 
underlying stellar populations, independent of what were obtained 
from the other wavelengths. Given the X-ray observations of NGC 1291, we
consider that the ring of this galaxy hosts a younger stellar 
population than the bulge, based on the arguments given below.
\begin{enumerate}

\item
There appears to be an overdensity of X-ray sources in the
ring of the galaxy.
The radial profile of the X-ray source surface density 
follows
the NUV surface brightness distribution in the ring 
instead of the $K$-band surface brightness distribution, suggesting that 
the ring XRBs are likely associated with a younger stellar population
(see Section \ref{radprofilesec}).

\item
The ring XRB population has in general a harder soft X-ray color than 
the bulge population, likely due to increased obscuration in the ring
associated with ongoing/recent star-forming activities 
(see Section \ref{xcolor}).

\item
The bulge and ring combined XLF can be well modeled with a synthetic XLF 
that is in general agreement with the PS modeling of typical old elliptical
galaxies. However, there appears to be an enhancement of 
super-Eddington XRBs, which could be associated a possible younger stellar    
population in the ring region (see Section \ref{modeling}).

\item
The XLF is an efficient probe of the relative age of stellar populations.
It has been reported that young elliptical galaxies
host a larger fraction of luminous LMXBs 
and the XLFs appear flatter than those of typical old elliptical galaxies 
\citep{Kim2010}. 
For NGC 1291, the slope of the broken power-law fit at the faint end
(Table~\ref{tbl-fit}) for the ring appears flatter than that for the 
bulge, albeit with a large uncertainty, suggesting that there are more
luminous X-ray sources in the ring. 
A comparison of the numbers of sources in given luminosity bins for the bulge
and ring populations is shown in Figure~\ref{scaleplot}a. 
The numbers of sources have been corrected for 
the detection incompleteness and background AGN contamination.
The fraction of relatively luminous sources 
($L_{\rm X}>5.5\times 10^{37}$~\lum)) is $30_{-7}^{+8}\%$ for the bulge,
and it is $54_{-11}^{+14}\%$ for the ring; the errors are Poissonian.

The LMXBs of the ring likely belong to two populations, 
a relatively old population that ages similarly
to the bulge population 
and a relatively young population that formed with the enhanced star formation.
To isolate this young population, we simply assume that 
the source number counts of the 
old population scale with the total $K$-band luminosity, and then the 
old population in the ring can be derived from the bulge source number
counts and the 
ratio between the $K$-band luminosities of the bulge and ring (a factor of 
3.6). We subtract this old population from the ring number counts
and the residual 
is displayed in Figure \ref{scaleplot}b. 
The fraction of relatively luminous sources
is $72_{-17}^{+22}\%$ for this remaining population in the ring.
The difference in the fraction of luminous sources between the bulge
and the ring appears more significant, and it is likely that 
the remaining XRB population is indeed younger.

\begin{figure*}
\centerline{
\includegraphics[scale=0.5]{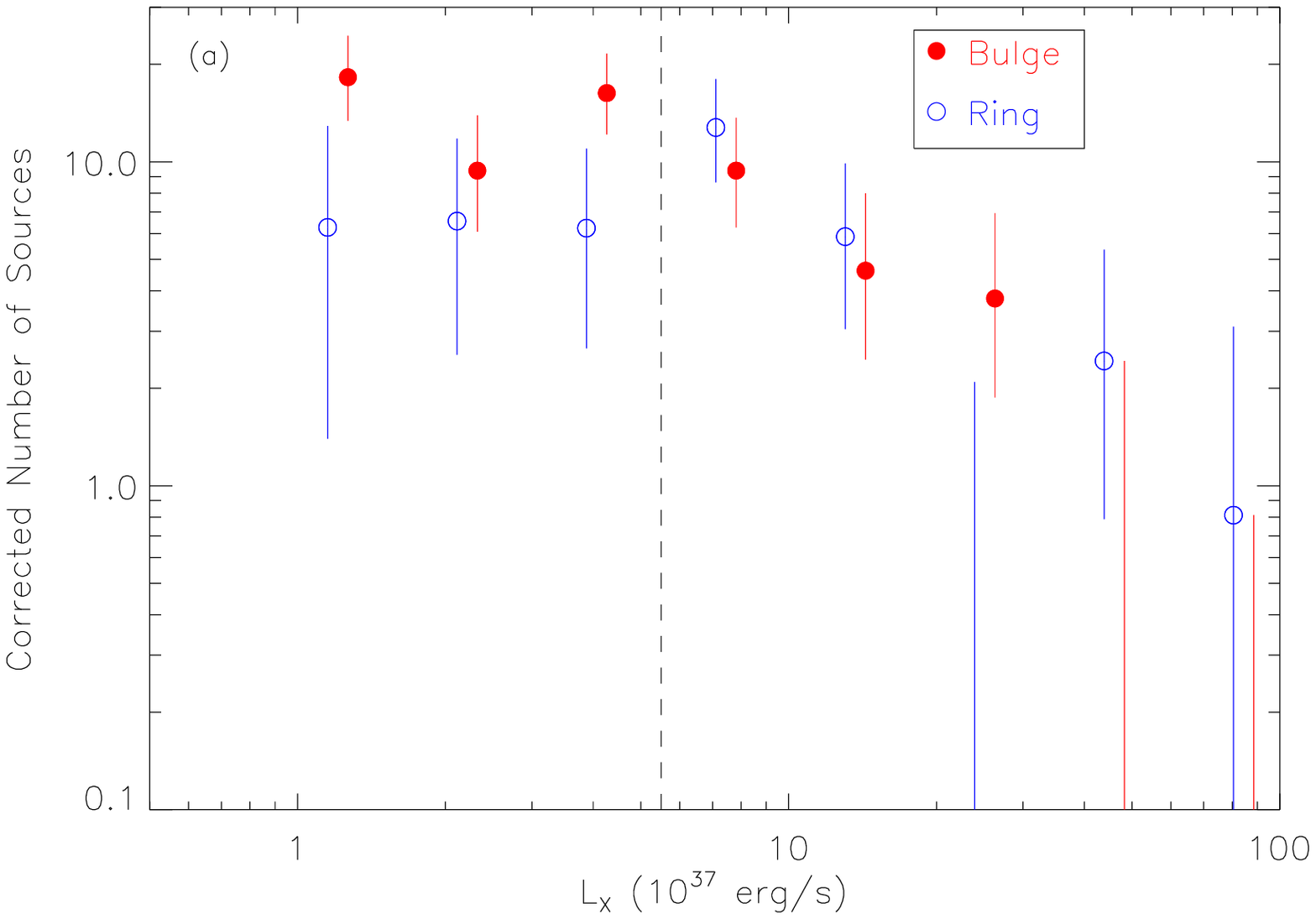}
\includegraphics[scale=0.5]{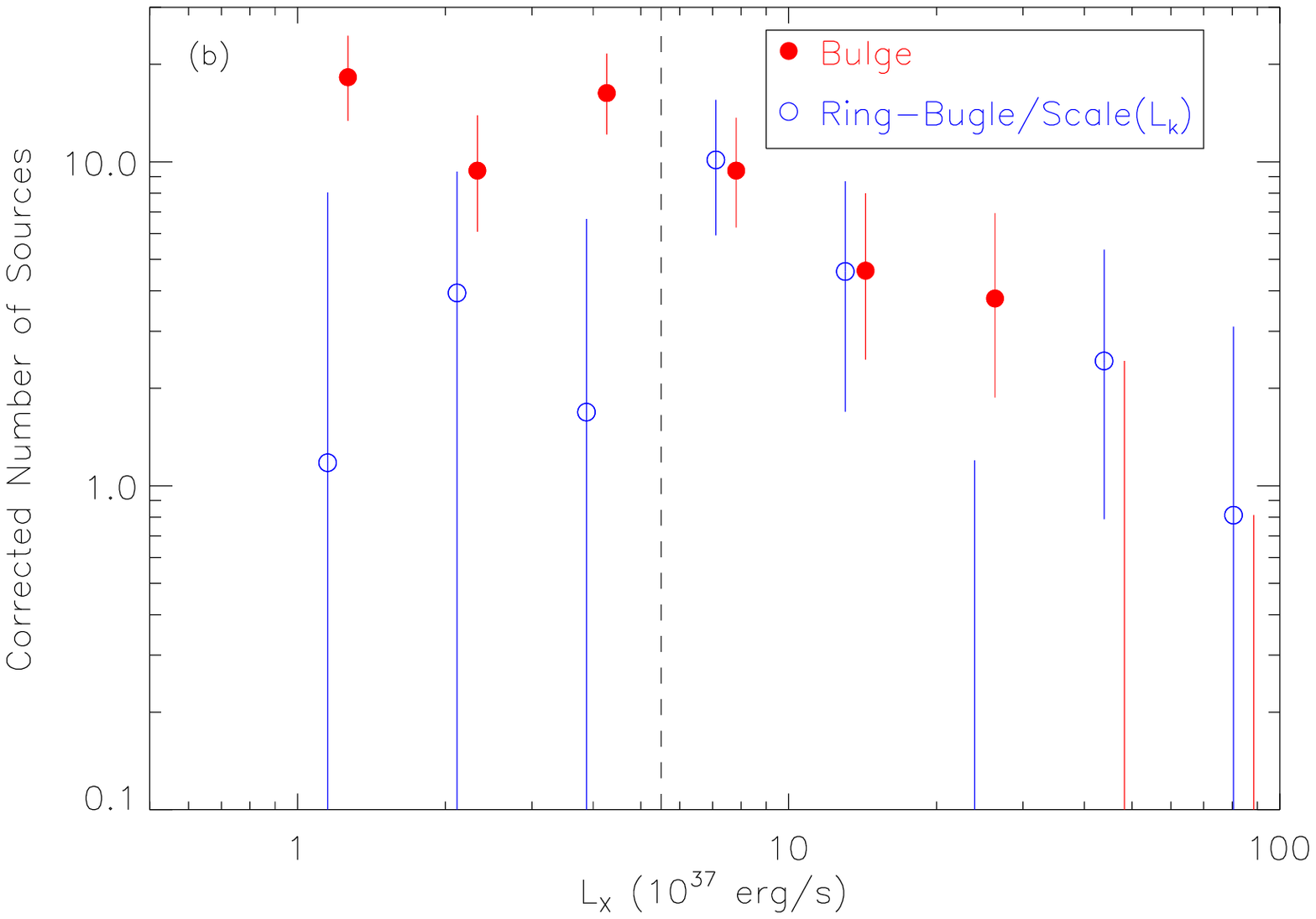}}
\figcaption{
Numbers of X-ray sources in given luminosity bins for the bulge,
ring [panel (a)], and bulge-subtracted ring [panel (b); see text]
populations. The numbers of sources have been corrected for
the detection incompleteness and background AGN contamination.
The luminosity bins were chosen to be the same for all the populations,
and the limiting luminosity is $\approx8.5\times10^{36}$~\lum. The 
luminosities for the bulge data points were shifted by a small amount for 
display purpose. For several 
cases with only error bars and no data points shown in the plot,
the numbers of sources are smaller than 0.1. The vertical dashed lines
mark the approximate breaks of the XLFs, and are used to separate 
relatively luminous and relatively faint sources.
(A color version of this figure is available in the online journal.)
\label{scaleplot}}
\end{figure*}

\end{enumerate}

\subsection{XRB Population of Galactic Rings}

As the two best studied galactic rings in the X-ray,
the rings of NGC 1291 and the Cartwheel galaxy differ significantly.
The XRB population in the ring of the Cartwheel galaxy
is dominated by ULXs; the XLF has a power-law slope of
$\approx1.6$ albeit at high luminosities ($L_{\rm X}\ga10^{39}$~\lum; 
\citealt{Wolter2004}), consistent with being dominated by
luminous young HMXBs. For NGC 1291, there are only two ULXs in the ring,
and the ring XLF shape and X-ray source colors suggest a LMXB dominated
system. The 
ring of the Cartwheel galaxy clearly has much stronger ongoing
star formation than the ring of NGC 1291 (total SFR of 
0.4~$M_{\sun}$~yr$^{-1}$ for NGC 1291 
compared
to 67.5~$M_{\sun}$~yr$^{-1}$ for the Cartwheel galaxy).

Such a discrepancy is probably caused by different formation and evolution
scenarios of
the galactic rings. The ring of the Cartwheel galaxy is likely
produced by a recent merger event, and the expanding density wave
forms the starburst ring. 
Collisional rings are generally short-lived with a lifetime of
the order of a few hundred
million years (see the review of \citealt{Moiseev2009}
and references therein).
The ring of NGC 1291 is probably created
by secular gas accumulation at the resonance orbit
due to the gravity torques of the bars. 
Enhanced star formation appeared
in the ring as gas was being gathered. However, this could a be a less
energetic and longer-term event than those appeared 
in collisional ring systems,
resulting in the relatively mild star formation observed today.
Adopting the \citet{Noll2009} star-formation history model (Section 1)
and assuming the ring hosts the young stellar population with an
e-folding timescale of 50 Myr, the SFR of NGC 1291 would be 
$\approx20$~$M_{\sun}$~yr$^{-1}$ 200 Myr ago, comparable to the current 
SFR of 
the Cartwheel galaxy.

\section{Summary}

We presented \chandra\ observations of the ring galaxy NGC 1291 and 
studied the XRB populations in the ring and bulge regions of the galaxy.
The main results are summarized below.

\begin{enumerate}
\item
We analyzed the four \chandra\ ACIS-S3 observations of NGC 1291, with a 
total exposure of $\approx180$~ks. We have detected 169 point sources
in the 0.3--8.0 keV band, 
75 of which are in the bulge and 71 in the ring. 
The full-band luminosities of the sources range from
$2.7\times10^{36}$ to $2.3\times10^{39}$~\lum.

\item
We studied basic properties of the X-ray sources. 
There appears to be an overdensity of X-ray sources in the
ring of the galaxy. 
The radial profile of the X-ray source surface density
resembles the NUV surface brightness distribution in the ring, 
suggesting that a significant fraction of 
the ring XRBs are associated with a relatively young 
stellar population. About 40\% of the sources in the bulge and 
25\% of the sources in the ring show $>3\sigma$ long-term variability in the
full-band count rate. 

\item
We studied the spectra of the nuclear source, which appears to be
a low-luminosity AGN with moderate obscuration, and there is also 
hot gas present 
in the nuclear region that produces soft excesses in the spectra. 
The spectra of individual observations
were well fit with an
absorbed power-law plus thermal plasma model.

\item
In the X-ray color-color plot, a significant
fraction of the bulge ($\approx75\%$) and ring ($\approx65\%$) 
sources with relatively small color errors are located in the 
region expected for LMXBs. On average, the ring XRB population
has a harder soft X-ray color, suggesting that the ring 
XRBs are more obscured, probably associated with ongoing/recent 
star-forming activities in the ring.

\item
We constructed full-band XLFs for the bulge and ring regions, taking into 
account the detection incompleteness and background AGN
contamination.
The resulting 90\% completeness limits are $\approx1.5\times10^{37}$
and
$\approx2.2\times10^{37}$~\lum\ for the bulge and ring populations, 
respectively. 
The ring XRB population is heavily contaminated
($\approx35\%$)
by background AGNs. Both XLFs can be fit with a broken power-law model.
The shapes are consistent with those expected for populations dominated by
LMXBs.
The ring XLF appears flatter than the bulge XLF, indicating 
that the ring hosts a larger fraction of luminous sources and is likely
associated with a younger stellar population.

\item
We successfully modeled the bulge and ring combined XLF with a PS model,
utilizing our current knowledge of the galaxy properties. 
The model is consistent with those for typical old elliptical
galaxies, suggesting that an old LMXB population is 
dominating the observed XLF. The relative enrichment of luminous sources 
of NGC 1291 can be attributed a younger XRB population 
originated from the ring star-forming region.

\item
We consider that the ring of NGC 1291 hosts a younger stellar
population than the bulge based on the X-ray studies
presented in this paper.

\end{enumerate}

~\\

This work is supported by NASA grant GO0-11104X. We acknowledge 
support from the CXC, which is operated by the 
Smithsonian Astrophysical 
Observatory (SAO) for and on behalf of NASA under Contract NAS8-03060.
S. Pellegrini acknowledges partial support from ASI/INAF grant I/009/10/0.
P. Tzanavaris acknowledges 
support through a NASA Postdoctoral Program Fellowship at
Goddard Space Flight Center, administered by Oak
Ridge Associated Universities through a contract with NASA.
We thank the referee for carefully
reviewing the manuscript and providing helpful comments.

\begin{deluxetable}{lcccl}
\tablewidth{0pt}

\tablecaption{\chandra\ Observations of NGC 1291}
\tablehead{ 
\colhead{Obs. ID}                                 &
\colhead{Start Date}                                 &
\colhead{Exp (ks)}                             &
\colhead{Cleaned Exp (ks)}                         &
\colhead{PI} \\
\colhead{(1)}         &
\colhead{(2)}         &
\colhead{(3)}         &
\colhead{(4)}         &
\colhead{(5)}             
}
\startdata
795 &2000 Jun 27 &39.7 & 37.6 & J.~A. Irwin\\
2059 &2000 Dec 7  &37.0 & 22.8 &A. H. Prestwich\\
11272 & 2010 May 4 &  70.0 & 69.1&G. Fabbiano\\
11273 &2010 May 8 & 50.0 & 49.1 & G. Fabbiano\\
\enddata
\tablecomments{Column 1: \chandra\ observation identification number. 
Column 3: Observation start data.
Column 3: Nominal exposure time. 
Column 4: Exposure time after removing background flares.
Column 5: Name of the Principal Investigator.}
\label{tbl-obs}
\end{deluxetable}

\begin{deluxetable}{lllcccccccc}
\tabletypesize{\scriptsize}
\tablewidth{0pt}    
\tablecaption{{\it Chandra} Point-Source Catalog}

\tablehead{
\colhead{} &
\multicolumn{2}{c}{X-ray Coordinates} &
\colhead{}                   &
\colhead{}                   &
\multicolumn{6}{c}{Counts}      \\
\\ \cline{2-3} \cline{6-11} \\
\colhead{XID}                    &
\colhead{$\alpha_{2000}$}       &
\colhead{$\delta_{2000}$}       &
\colhead{Off-Axis}       &
\colhead{$d_{\rm rad}$}       &
\colhead{FB}          &
\colhead{FB Low Err}          &
\colhead{FB Up Err}          &
\colhead{SB}          &
\colhead{SB Low Err}          &
\colhead{SB Up Err}          \\
\colhead{(1)}         &
\colhead{(2)}         &
\colhead{(3)}         &
\colhead{(4)}         &
\colhead{(5)}         &
\colhead{(6)}         &
\colhead{(7)}         &
\colhead{(8)}         &
\colhead{(9)}        &
\colhead{(10)}      &
\colhead{(11)}      
}
\startdata
      1 \dotfill \ldots   &03 16 47.17 &$-$41 08 15.2& 4.6&6.2&29.9&     5.9&   7.2&    19.2&    4.8&    6.2\\
 2 \dotfill \ldots   & 03 16 47.62&  $-$41 05 34.2 & 4.2&  5.9 &  144.8 &   13.0 &   14.2 &  106.9 &   11.3 &   12.6\\
   3  \dotfill \ldots   & 03 16 48.09 & $-$41 04 10.1 & 4.6&  6.2 &    6.6 &    2.8  &   4.2 &   10.8 &   $-$1.0 &   $-$1.0\\  
\enddata
\tablecomments{
Units of right
ascension are hours, minutes, and seconds, and units of declination are
degrees, arcminutes, and arcseconds.
(This table is available in its entirety in a machine-readable 
form in the online journal. A portion is shown here for guidance regarding its form
and content. The full table contains 34 columns of
information on the 169 \hbox{X-ray} sources.)}
\label{tbl-mcat}

\end{deluxetable}

\begin{deluxetable}{lcccc}
%\tabletypesize{\small} 
%\rotate
\tablewidth{0pt}

\tablecaption{Best-Fit Parameters for the Nuclear Source}
\tablehead{
\colhead{Observation}             &
\colhead{$\chi^2/$dof}   &
\colhead{$\Gamma$}   &
\colhead{$N_{\rm H,2}$}   & 
\colhead{$L_{2\textrm{--}8}$} \\
\colhead{}   &
\colhead{}   &
\colhead{}   &
\colhead{($\times10^{22}$~cm$^{-2}$)}   &
\colhead{($\times10^{39}$~\lum)} \\
\colhead{(1)}   &
\colhead{(2)}   &
\colhead{(3)}   &
\colhead{(4)}   &
\colhead{(5)}   \\
}

\startdata
795&0.84&$1.6^{+0.6}_{-0.5}$&   $2.0^{+0.4}_{-0.4}$
& $2.57^{+0.08}_{-0.44}$ \\
2059&0.93&$2.6^{+1.4}_{-1.2}$&  $3.3^{+1.7}_{-1.3}$ 
& $1.28^{+0.07}_{-0.91}$ \\
11272&1.21&$2.1^{+0.4}_{-0.3}$&  $1.8^{+0.5}_{-0.4}$
& $1.29^{+0.04}_{-0.13}$ \\
11273&1.11&$1.8^{+0.5}_{-0.5}$&  $1.8^{+0.8}_{-0.6}$
& $1.27^{+0.04}_{-0.23}$ \\
\enddata
\tablecomments{The spectra were fit with an 
absorbed power law plus thermal plasma
model ({\sc wabs1*apec+wabs2*pow}). 
$N_{\rm H,2}$ 
and $\Gamma$ are the absorption column density and photon index 
for the power-law component, respectively.
The temperature ($0.15$~keV) 
and column density ($N_{\rm H,1}=7.6\times
10^{21}$~cm$^{-2}$) of the thermal component were
fixed at the values derived from the stacked spectrum.
The observed 2.0--8.0 keV luminosities are
dominated by the
power-law component.
}
\label{specfit}
\end{deluxetable}

\begin{deluxetable}{lcccccccccc}
\tabletypesize{\small} 
%\rotate
\tablewidth{0pt}

\tablecaption{Best-Fit Parameters for the X-Ray Luminosity Functions}
\tablehead{
\colhead{Region} &
\multicolumn{3}{c}{Single Power-law}     &
\colhead{ }                                 &
\multicolumn{5}{c}{Broken Power-law}      \\
\\ \cline{2-4} \cline{6-10} \\
\colhead{ }                                 &
\colhead{$S_{\rm p}$}                             &
\colhead{$\alpha$}                             &
\colhead{$A1$}            &
\colhead{ }                                 &
\colhead{$S_{\rm bp}$}                             &
\colhead{$\alpha_1$}                                 &
\colhead{$\alpha_2$}                             &
\colhead{$L_{\rm b}~(\times10^{37}$~\lum)}              &
\colhead{$A2$}           
}
\startdata
Bulge & 143 & $1.6^{+0.1}_{-0.1}$ & $36^{+7}_{-6}$&&138&$1.1^{+0.3}_{-0.6}$& $2.0^{+0.5}_{-0.3}$ & $4.9^{+4.9}_{-2.1}$
& $26^{+8}_{-9}$ \\
Ring & 159 &$1.4^{+0.1}_{-0.1}$ & $18^{+7}_{-6}$ & &153&$0.4^{+0.5}_{-0.7}$ & $2.1^{+0.5}_{-0.3}$ & $6.4^{+3.1}_{-1.1}$
& $6^{+7}_{-4}$ \\

\enddata
\tablecomments{$S_{\rm p}$ and $S_{\rm bp}$ are the Cash fit statistic values
for the power-law and broken power-law fit, respectively. See Equations (2) and
(3) for the definition of the other parameters.}
\label{tbl-fit}
\end{deluxetable}

\begin{deluxetable}{lcc}
\tablecolumns{3}
%\tabletypesize{\scriptsize}
\tablewidth{0pt}
\tablecaption{Model Parameters for NGC1291
\label{modelparam}}
\tablehead{ \colhead{Parameter} &
     \colhead{Notation} &
     \colhead{Value}
     }
\startdata
Star Formation History                   &           & \citet{Noll2009} \\
Metallicity           & $Z$           & 0.02 \\
Total Stellar Mass        & $M_{*}$ &  $ 1.5 \times 10^{11}\, \rm M{\odot}$ \\
Binary Fraction  & $F_{\rm bin}$ & 50\% \\ 
IMF power-law slope &      & -2.7 or -2.35 \\
CE Efficiency & $\alpha_{\rm CE}$  &  0.1, 0.2, 0.3, or 0.5 \\
Stellar wind strength & $\eta_{\rm wind}$  &  0.25, 1.0, or 2.0 \\
CE during HG & & Yes or No \\
Magnetic Braking &              & \citet{IT2003}

\enddata
\label{model_param}
\end{deluxetable}

\end{document}